\documentstyle[12pt,epsf,graphics]{article}
\def\be{\begin{equation}}
\def\ee{\end{equation}}
\def\beqna{\begin{eqnarray}}
\def\eeqna{\end{eqnarray}}
\def\bear{\begin{eqnarray}}
\def\eear{\end{eqnarray}}
\def\bearst{\begin{eqnarray*}}
\def\eearst{\end{eqnarray*}}
\begin{document}

\begin{center}
{\large\bf Shortcuts in  Domain Walls and the Horizon Problem}
\end{center}
\vspace{1ex}
\centerline{\large Elcio Abdalla and Bertha Cuadros-Melgar}
\begin{center}
{Instituto de Fisica, Universidade de S\~ao Paulo\\
C.P.66.318, CEP 05315-970, S\~ao Paulo, Brazil.}
\end{center}
\vspace{6ex}

\begin{abstract}

We consider a dynamical membrane world in a space-time with 
scalar bulk matter described by domain walls. Using the solutions to
Einstein field equations and Israel conditions we investigate the
possibility of having shortcuts for gravitons leaving the wall and
returning subsequently. As it turns out, they usually appear
under mild conditions.

In the comparison with  photons following a geodesic inside the brane, 
we verify that shortcuts exist. For some Universes they are small, but
there are cases where shortcuts are effective. In these cases we
expect them to play a significant role in the solution of the horizon
problem.

\end{abstract}

\section{Introduction}
Although standard model
of particle physics has been established as the uncontested theory of all
interactions down to distances of $10^{-17}$m, there are good reasons to
believe that there is a new physics arising soon at the experimental
level \cite{veneziano}. On the other hand, string theory provides an excelent 
background to solve long standing problems of theoretical high energy
physics. It is by now a widespread idea
that M-theory \cite{polchinski} is a reasonable description
of our Universe. In the field theory limit, it is described by a solution 
of the (eventually 11-dimensional) Einstein equations with a cosmological 
constant by means of a four dimensional membrane. In this picture only 
gravity survives in the extra dimensions, while the remaining matter and 
gauge interactions are typically four dimensional.

In this picture, there is a
possibility that  gravitational fields, while propagating out of the brane
speed up, reaching farther distances in smaller time as compared to
light propagating inside the brane, a scenario that for a resident of the
brane (as ourselves) implies shortcuts \cite{Abdalla2}---\cite{Chung}.

This possibility implies that we can have alternatives to the inflationary
scenario in order to explain the homogeneity problem in cosmology. 
Recently, this scenario has been proposed as an actually realizable
possibility \cite{stojkovic}---\cite{moffat}. In \cite{abdacasali} it
has been shown that in some scenarios shortcuts are very difficult to
be detected today because of 
the extremely short delay of the photon as compared to the graviton coming
from the same source. However, in cases where before nucleosynthesis
delays are large enough to imply thermalization of the whole universe, a 
possible solution to the homogeneity problem \cite{abdacasali},
alternative \cite{stojkovic,freese,moffat} to inflation \cite{turner}
may exist based 
on these shortcuts. In the case of domain walls shortcuts are common
and a possibility of solving the homogeneity problem may be under way,
though further work is certainly necessary.

Recently it has been proved \cite{transdimen} that  brane 
Universes  can provide a means for finding relics of the
higher dimensions in the cosmic microwave background as well. Thus it is
worthwhile further pursuing  brane models as useful tools to understand 
the physics of strings and M-theory \cite{generalbranes}, which proves
them as very important instances to try a better insight of
the Universe and its properties.

At last, we should stress that this is a toy model where the universe
is replaced by a domain wall. The more realistic case deserves
attention as well (see also \cite{abdacasali}).

\section{The General Setup}
We consider a scenario described by the gravitational
action in a D-dimensional bulk with a scalar field, a bulk dilaton, 
a domain wall potential and a Gibbons-Hawking term \cite{gibbonshawking},
\be\label{action}
S = \int_{bulk} d^D x \sqrt{-g} \left( {1 \over 2} R - {1 \over 2} (\partial
  \phi )^2 -V(\phi) \right) - \int_{dw} d^{D-1} x \sqrt{-h} ([K] +
  \hat V (\phi) ) \, ,
\ee
where $\phi$ is the bulk dilaton, $K$ is the extrinsic curvature, $V(\phi)$ 
and $\hat V (\phi)$ are bulk and domain wall potentials respectively,
and $g$ and $h$ denote the bulk and domain wall metrics. 
The potentials are here considered to be of the Liouville type:
\begin{eqnarray}
V(\phi) &=& V_0 e^{\beta \phi} \, , \label{vbulk} \\
\hat V (\phi) &=& \hat V_0 e ^{\alpha \phi} \, . \label{vbrane}
\end{eqnarray}

We consider the bulk metric as being static and invariant under rotation,
being given by 
\begin{equation}\label{metric}
ds^2 = -U(r) dt^2 + U(r)^{-1} dr^2 + R(r)^2 d\Omega_k ^2 \, ,
\end{equation}
where $d\Omega_k ^2$ is the line element on a $D-2$ dimensional space
of constant curvature depending on a parameter $k$. Such a metric is 
supposed to have a 
mirror symmetry $Z_2$ with respect to the domain wall. Such a symmetry 
will be used in order to impose the Israel conditions \cite{israel}. 
In fact, the variation of the total action (\ref{action}) including the
Gibbons-Hawking term leads directly to the Israel conditions which in 
view of the $Z_2$ symmetry become
\begin{equation}\label{israel}
K_{MN} = - {1 \over {2(D-2)}} \hat V (\phi) h_{MN} \quad .
\end{equation}
The extrinsic curvature can be computed as
\begin{equation}\label{excurv}
K_{MN} = h^P _M h^Q _N \bigtriangledown_{_P} n_Q \quad ,
\end{equation}
where the unit normal, which points into $r<r(t)$, is
\begin{equation}\label{normal}
n_M = {1 \over \sqrt{U-{{\dot r^2}\over U}}} (\dot r \, , \, -1 \, ,
\, 0 \, ... \, , \, 0) \, .
\end{equation}
Here a dot means derivative with respect to the bulk time $t$.

The $ij$ component of (\ref{israel}) can be written as
\begin{equation}\label{ij}
{{R'}\over R} = {{\hat V(\phi)} \over {2(D-2)U}} \sqrt{U-{{\dot
r^2}\over U}} \; ,
\end{equation}
while the $00$ component is
\begin{equation}\label{00}
\left({{R'}\over R} \right) ^{-1} \left( {{R'} \over R} \right)
^\prime = {{\hat V'(\phi)} \over {\hat V(\phi)}} - {{R'}\over R} \; .
\end{equation}
Here a prime denotes derivative with respect to the radial extra
coordinate $r$.

The equation of motion for the dilaton obtained from the action 
(\ref{action}), together with  (\ref{00}), can be simultaneously solved
with the Ansatz (\ref{metric}), leading to \cite{chre}
\begin{eqnarray}
\phi(r) &=& \phi_\star - {{\alpha (D-2)} \over {\alpha^2 (D-2) + 1}}
\log r \, , \label{phi} \\
R(r) &=& (\alpha^2(D-2) +1) C \hat V_0 e^{\alpha \phi_\star}
r^{1\over {\alpha^2(D-2)+1}} \, , \label{R}
\end{eqnarray}
where $\phi_\star$ and $C$ are arbitrary integration constants. 

The motion of the domain wall is governed by the $ij$ component of the
Israel conditions (\ref{ij}). That equation can be written in terms of
the brane proper time $\tau$ as 
\begin{equation}\label{dwmotion}
{1\over 2} \left({{dR}\over{d\tau}} \right) ^2 + F(R) =0 \; .
\end{equation}

The induced metric on the domain wall is Friedmann-Robertson-Walker 
and (\ref{dwmotion})
describes the evolution of the scale factor $R(\tau)$. This equation
is the same as that one for a particle of unit mass and zero energy
rolling in a potential $F(R)$ given by
\begin{equation}\label{potential}
F(R) = {1 \over 2} U R'^2 - {1 \over {8(D-2)^2}} \hat V^2 R^2 \; .
\end{equation}
Notice that the solution only exists when $F(R) \leq 0$.

From the induced domain wall metric we find the relations between the time
parameter on the brane ($\tau$) and in the bulk ($t$) as given by
\bearst 
dt = {\sqrt{U+\left({{dr}\over{d\tau}}\right)^2}\over U} d\tau \; ,
\eearst 
so that
\begin{equation}\label{eq3}
\dot r \equiv {{dr}\over {dt}} = {{dr}\over {d\tau}} {{d\tau}\over {dt}} =
{{dr}\over {d\tau}} {U \over \sqrt{U+ \left({{dr}\over{d\tau}
}\right)^2}} \; ,
\end{equation}
where ${{dr}\over{d\tau}} = {{dR}\over{d\tau}} \left({{dR}\over
{dr}}\right)^{-1} $ can be obtained from (\ref{dwmotion}). Equation
(\ref{eq3}) describes the motion of a domain wall in the static
background as seen by an observer in the bulk.

Consider two points on the brane. In general, there are more than
one null geodesic connecting them in the D-dimensional spacetime. The
trajectories of photons must be on the brane and those of gravitons
may be outside. We consider the shortest path for both photons and gravitons.
For the latter, the geodesic equation is the same as the one considered in 
\cite{Abdalla2}, since the bulk metric is static:
\begin{equation}\label{geo}
\ddot r_g + \left( {1 \over r_g} - {3 \over 2} {{U'}\over U}\right) \dot
r_g^2 + {1\over 2} U \, U' - {U^2 \over r_g} = 0 \; .
\end{equation}
Again a dot means derivative with respect to the bulk time $t$.

The solutions of (\ref{eq3}) and (\ref{geo}) in terms of the bulk
proper time $t$ were obtained by means of a {\bf MAPLE} program. Now we
discuss the possibility of shortcuts in the cases of the various 
solutions describing different Universes defined by the domain wall solution.

\section{Solutions of the Geodesic Equation}
\subsection{Type I Solutions} 
We define the type I brane solutions as those for which  $\alpha=\beta=0$.
Consequently, the potentials become cosmological constants. The solution 
also has a constant dilaton $\phi=\phi_0$. A simple rescaling in the 
metric leads us to 
\begin{equation}\label{metric1}
ds^2 = -U(R)dt^2 + U(R)^{-1} dR^2 +R^2 d\Omega_k ^2 \; ,
\end{equation}
with
\begin{equation}\label{u1}
U(R) = k -2MR^{-(D-3)} - {{2V_0} \over {(D-1)(D-2)}} R^2 \; ,
\end{equation}
which corresponds to  a topological black hole solution in D dimensions 
with a cosmological constant.

As discussed in Ref. \cite{chre}, if the domain wall has positive energy 
density ($\hat V_0 >0$), the relevant part of the bulk spacetime is 
$R<R(\tau)$, which is the region containing the singularity. If it
has negative energy density ($\hat V_0 <0$), the relevant part is
$R>R(\tau)$, which is non-singular unless the wall reaches $R=0$.

\begin{figure*}[htb!]
\begin{center}
\leavevmode
\begin{eqnarray}
\epsfxsize= 5.5truecm\rotatebox{-90}
{\epsfbox{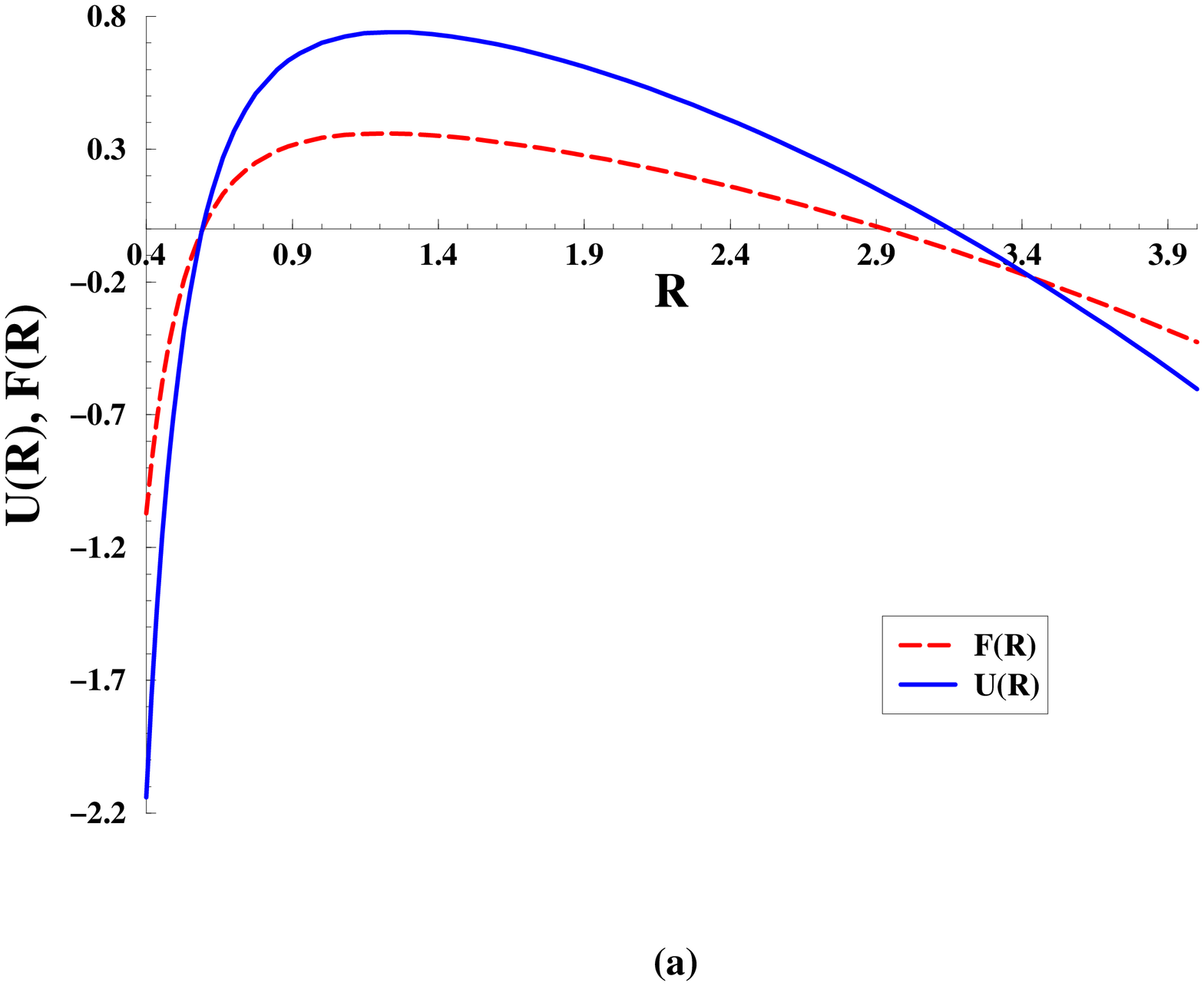}}\nonumber
\epsfxsize= 5.5truecm\rotatebox{-90}
{\epsfbox{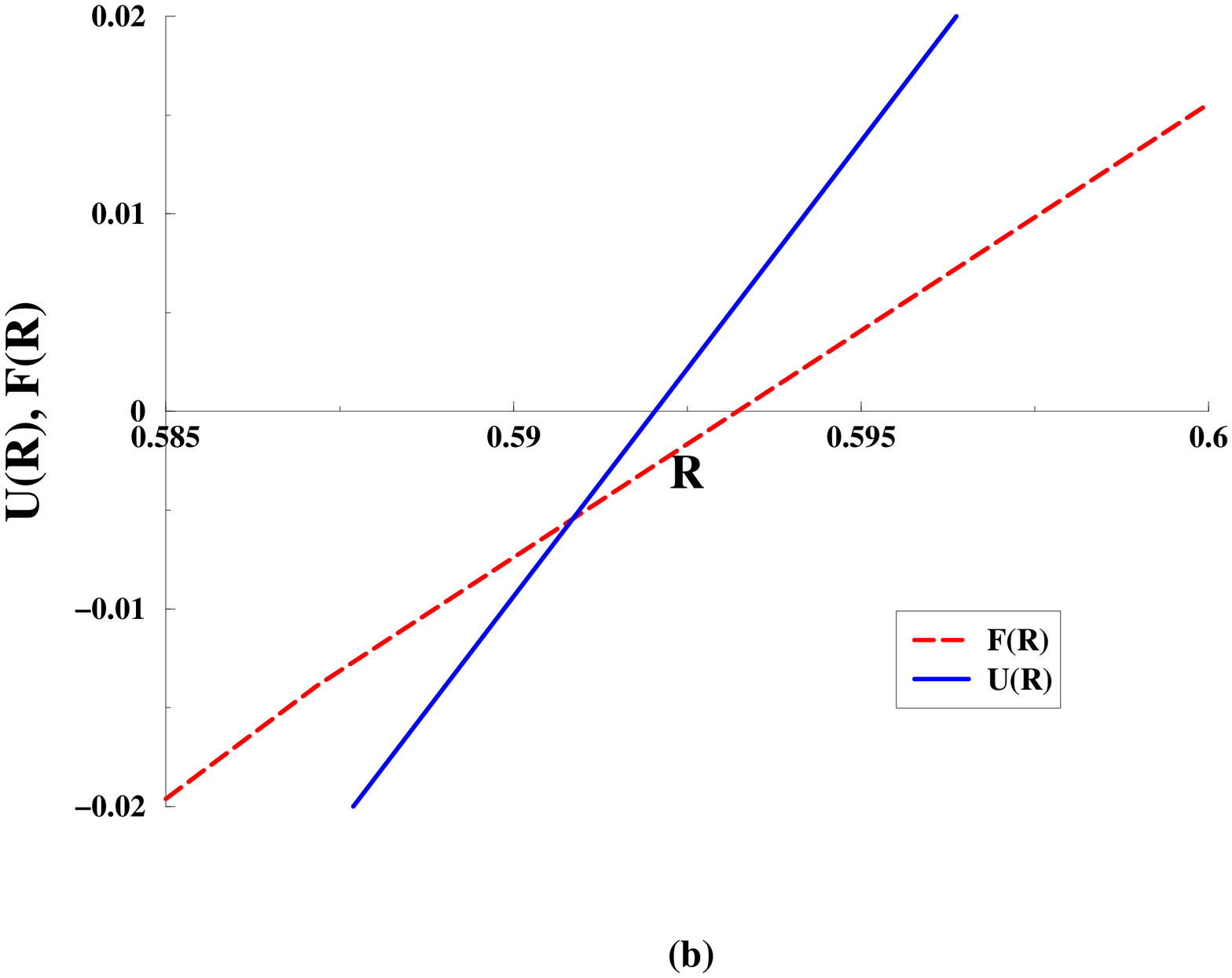}}\nonumber
\end{eqnarray}
\caption{(a) $U(R)$  and $F(R)$  for type I solutions with
$M=1/10$, $V_0=1$ and $\hat V_0=\pm 1$, (b) Zoom of the event horizon region.} 
\label{T1-U}
\end{center}
\end{figure*}
\begin{figure*}[htb!]
\begin{center}
\leavevmode
\begin{eqnarray}
\epsfxsize= 5.5truecm\rotatebox{-90}
{\epsfbox{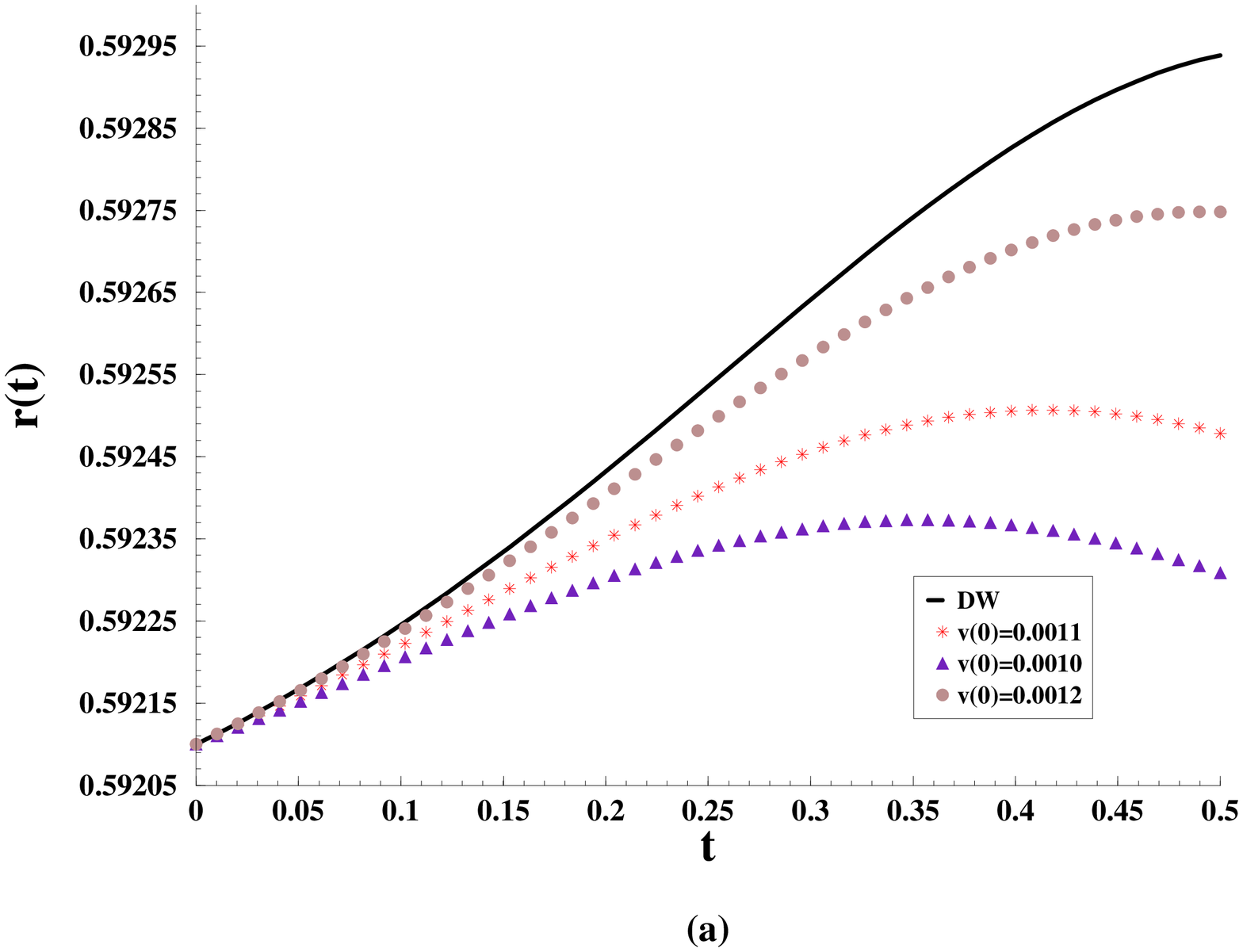}}\nonumber
\epsfxsize= 5.5truecm\rotatebox{-90}
{\epsfbox{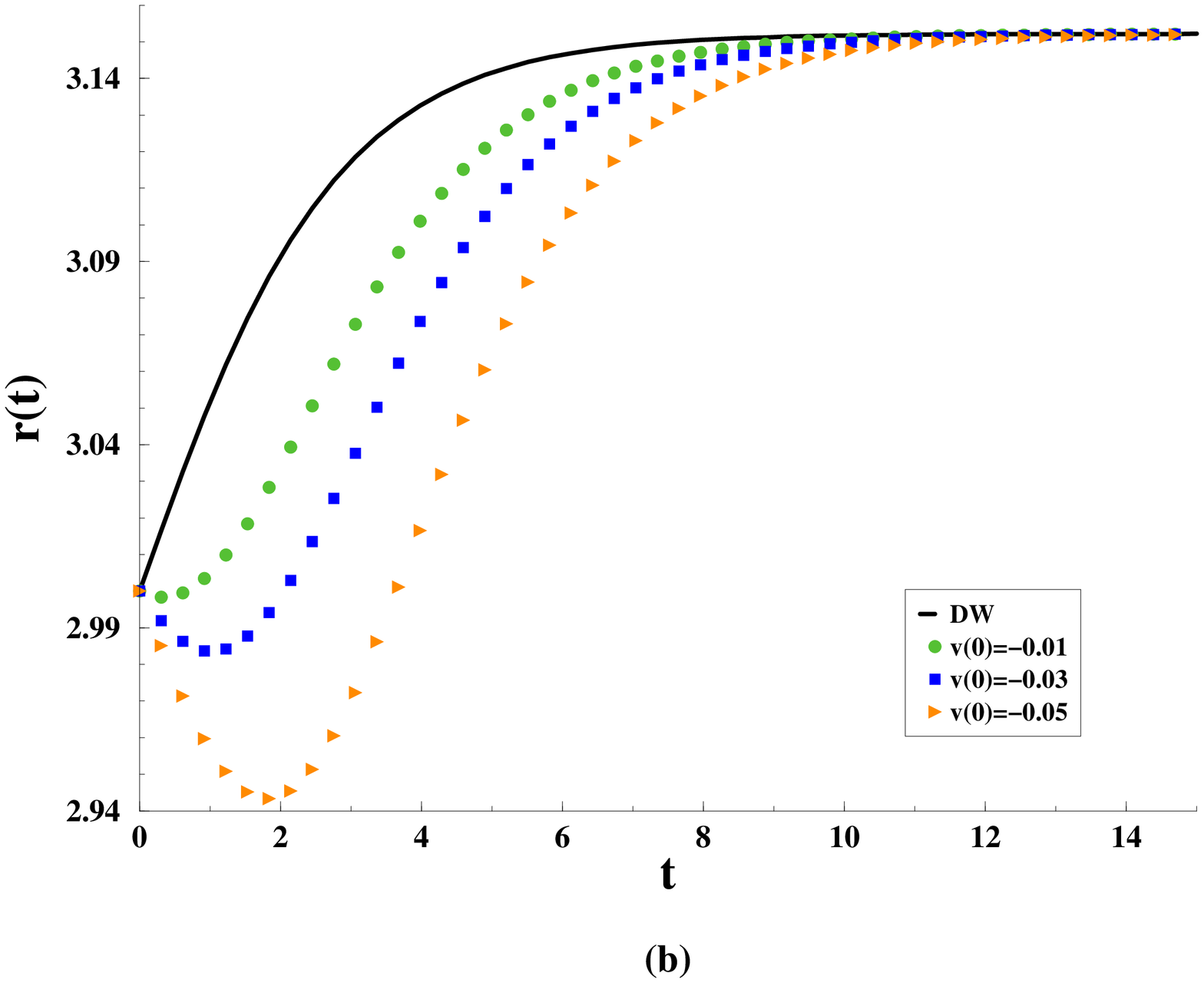}}\nonumber
\end{eqnarray}
\caption{Domain wall motion  and geodesics for type I solutions with
$M=1/10$, $V_0=1$ and $\hat V_0=1$ in (a) region I and, (b) region II.}
\label{T1-braneb}
\end{center}
\end{figure*}
\begin{figure*}[htb!]
\begin{center}
\leavevmode
\begin{eqnarray}
\epsfxsize= 5.5truecm\rotatebox{-90}
{\epsfbox{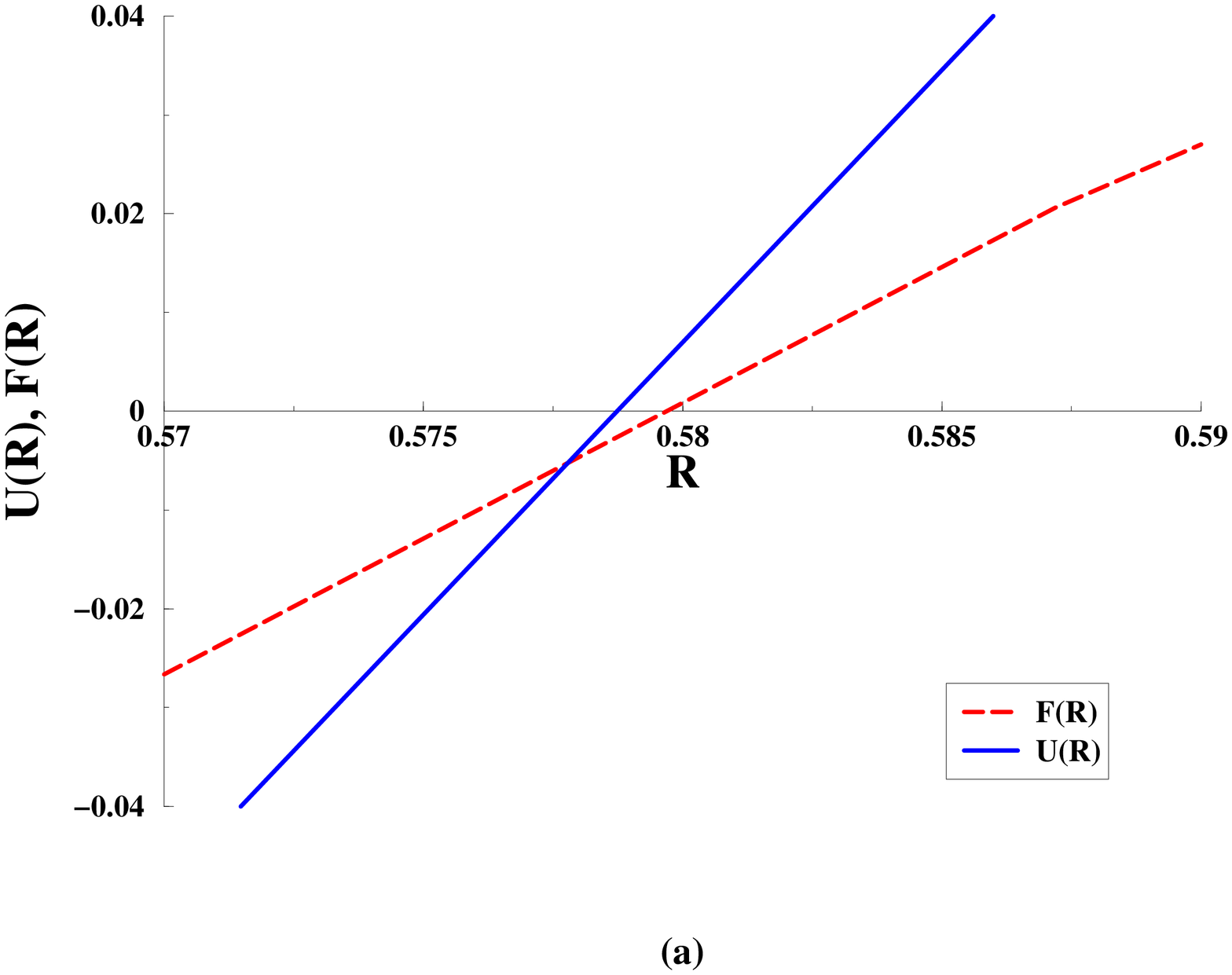}}\nonumber
\epsfxsize= 5.5truecm\rotatebox{-90}
{\epsfbox{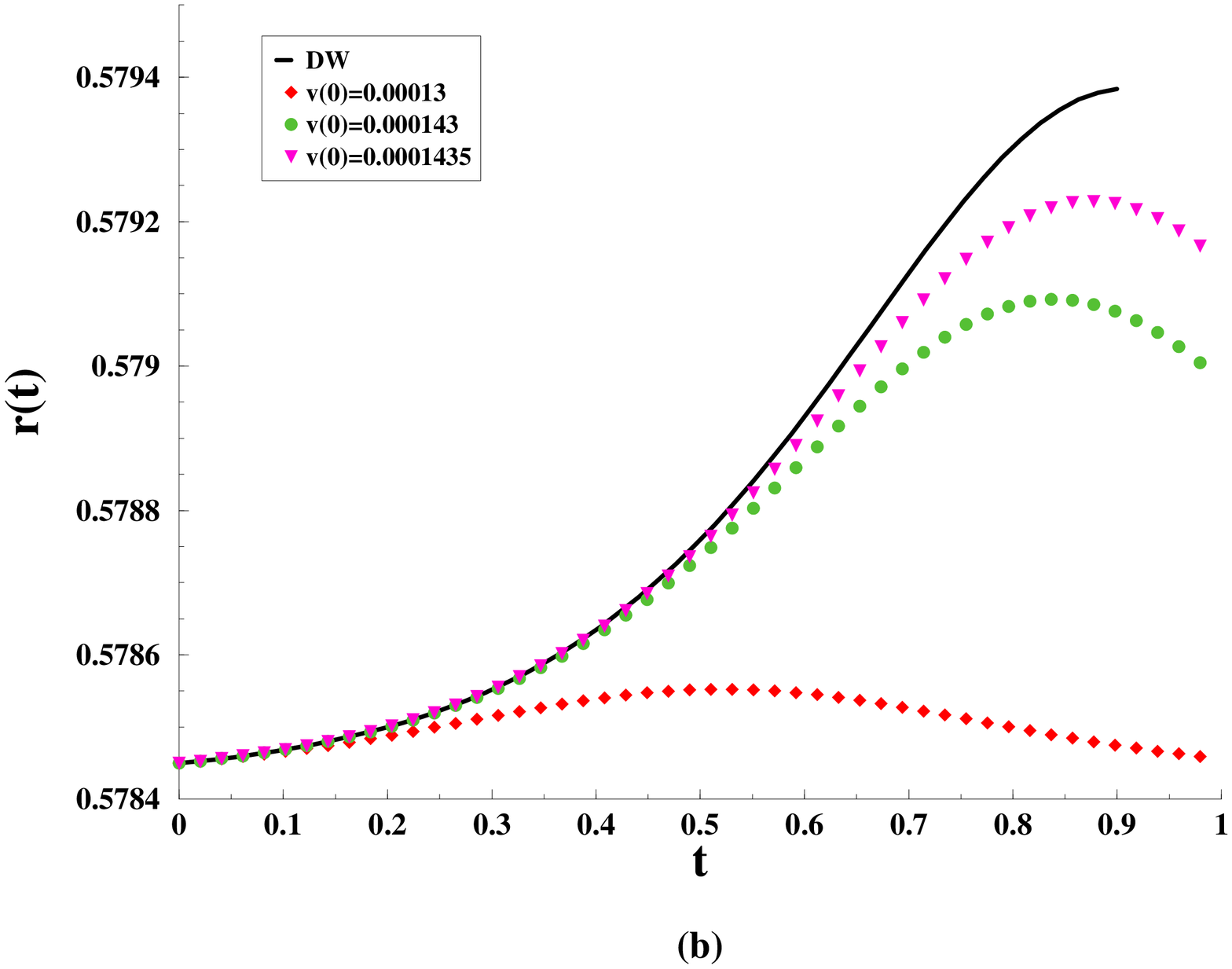}}\nonumber
\end{eqnarray}
\caption{
(a) Zoom of the region where (\ref{condition1}) holds from the graph of
$U(R)$ and $F(R)$ with $\hat \Lambda <0$ and $M>0$ for type I 
solutions, (b) Domain wall motion  and geodesics for type I solutions
with $M=1/10$, $V_0=-1$ and $\hat V_0=1$.}
\label{T1-U3}
\end{center}
\end{figure*}
\begin{figure*}[htb!]
\begin{center}
\leavevmode
\begin{eqnarray}
\epsfxsize= 5.5truecm\rotatebox{-90}
{\epsfbox{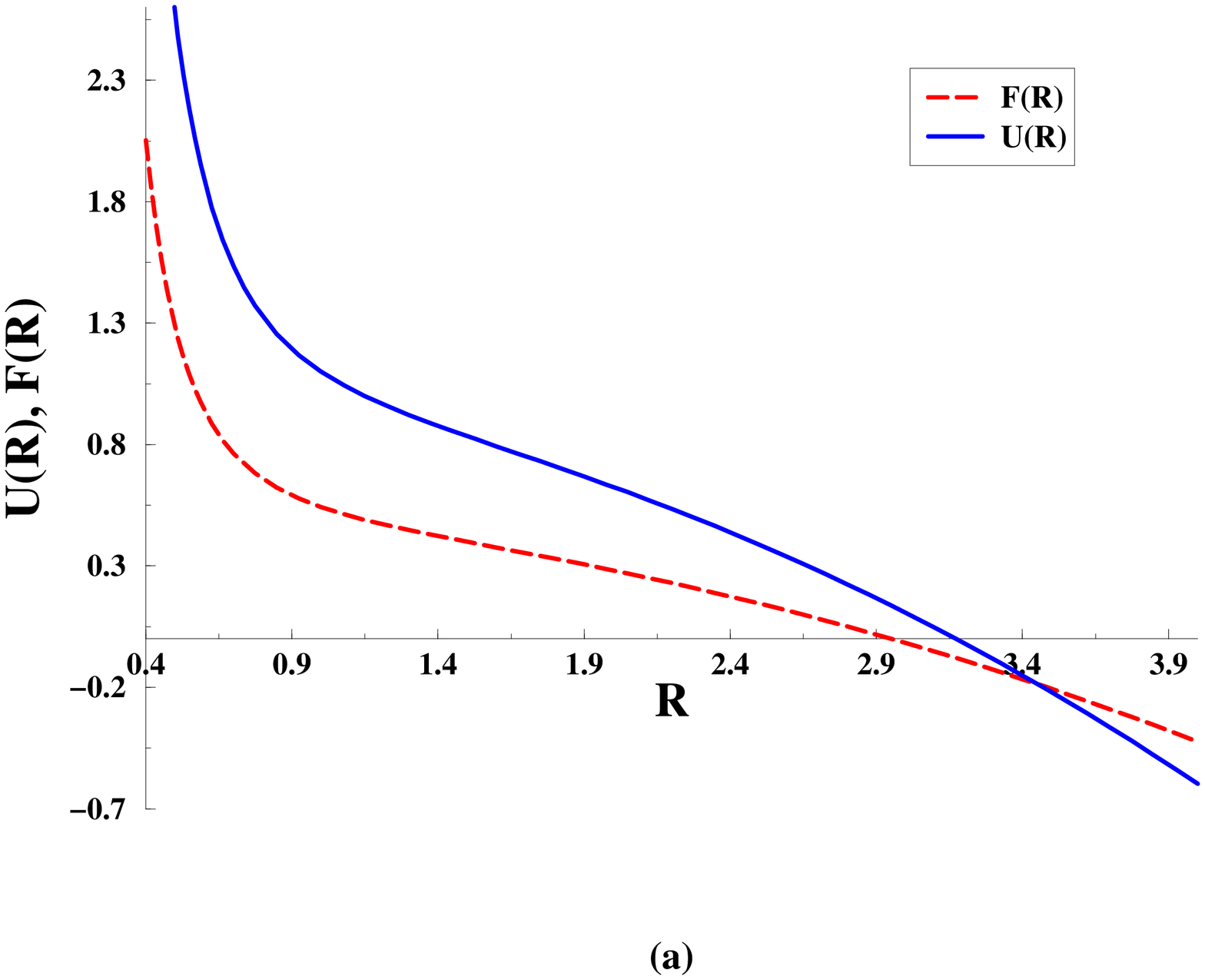}}\nonumber
\epsfxsize= 5.5truecm\rotatebox{-90}
{\epsfbox{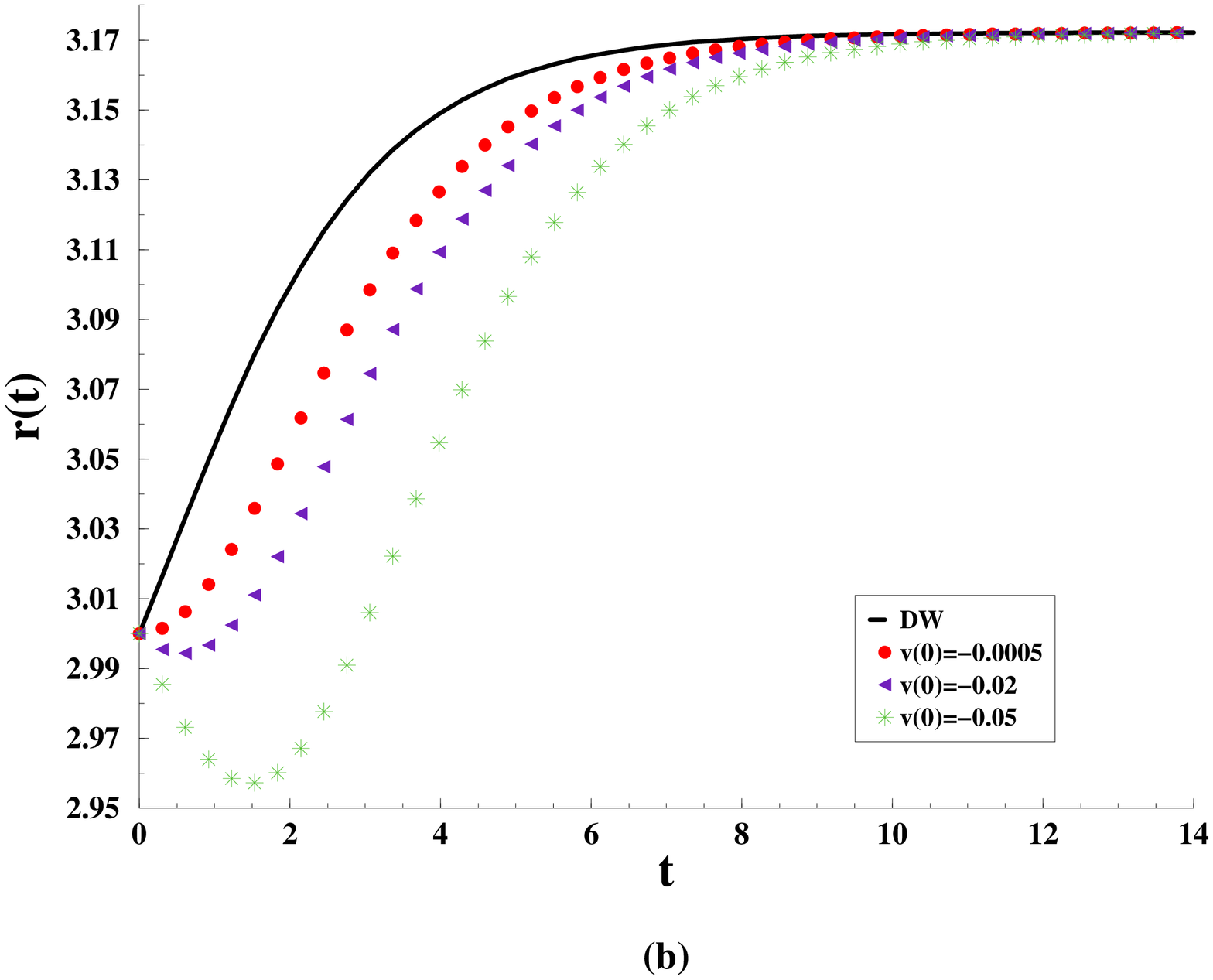}}\nonumber
\end{eqnarray}
\caption{(a) $U(R)$  and $F(R)$  with $\hat \Lambda >0$
and $M<0$ for type I 
solutions, (b) Domain wall motion and geodesics for $M=-1/10$, $V_0=1$
and $\hat V_0=1$.}
\label{T1-U2}
\end{center}
\end{figure*}
\begin{figure*}[htb!]
\begin{center}
\leavevmode
\begin{eqnarray}
\epsfxsize= 5.5truecm\rotatebox{-90}
{\epsfbox{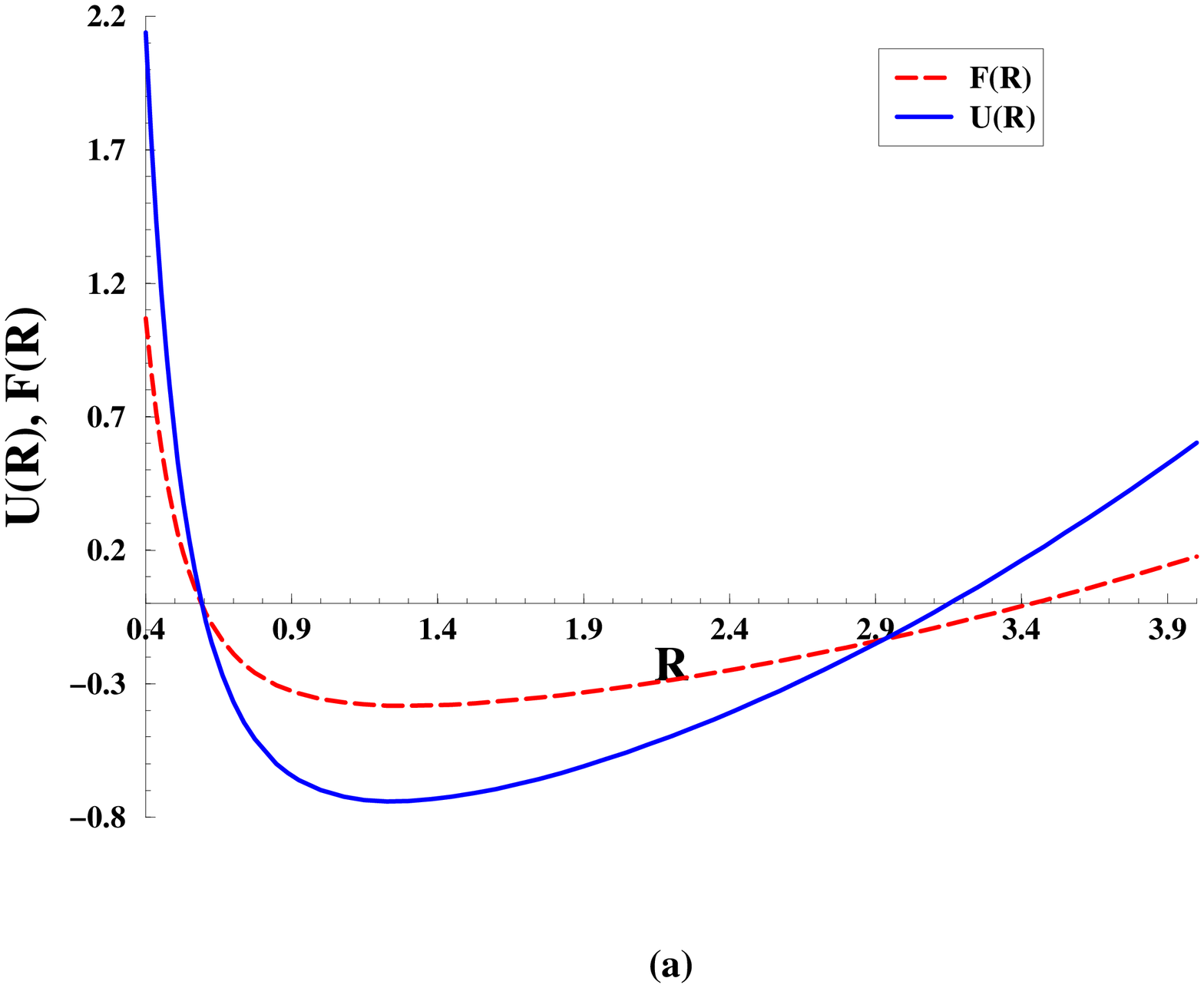}}\nonumber
\epsfxsize= 5.5truecm\rotatebox{-90}
{\epsfbox{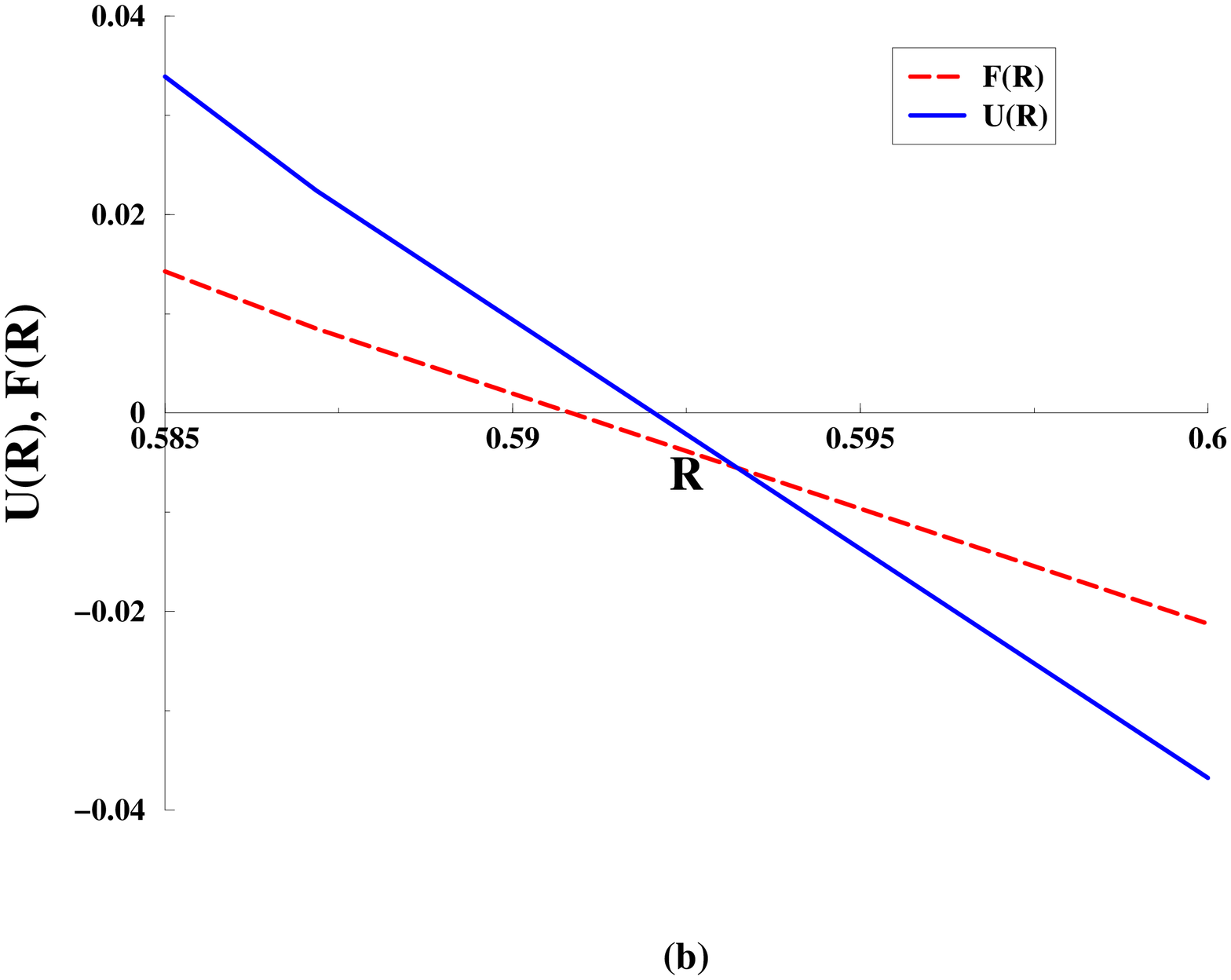}}\nonumber
\end{eqnarray}
\caption{(a) $U(R)$  and $F(R)$  with $\hat \Lambda <0$
and $M=-1/10$ for type I 
solutions, (b) Zoom of the event horizon region.}
\label{T1-U4}
\end{center}
\end{figure*}

The potential $F(R)$ ruling the evolution of the scale factor is
\begin{equation}\label{f1}
F(R)= {k \over 2} - MR^{-(D-3)} - \hat\Lambda R^2 \, ,
\end{equation}
where the effective cosmological constant on the domain wall is given by
\begin{equation}\label{lambda1}
\hat\Lambda = {1\over {D-2}} \left[ {{V_0}\over{D-1}} + {{\hat V_0 ^2}
    \over {8(D-2)}} \right] \, .
\end{equation}

We shall analyze each of the four cases presented in \cite{chre}.
As we have previously stated, the equation of motion (\ref{dwmotion})
has a solution only when $F(R) \leq 0$. This is automatic only if $U(R)<0$,
i.e. if $r$ is a time coordinate; therefore, we look for solutions with 
$U(R)>0$. In fact, both conditions,
\begin{equation}\label{condition1}
F(R) \leq 0 \qquad \hbox{and} \qquad U(R)>0 \, , 
\end{equation}
can coexist in some cases as we will see in what follows. In order to
ilustrate the following examples we have chosen $D=6$ dimensions.

\subsubsection{$\hat \Lambda>0$, $M>0$}
From the graph of $U(R)$ (see Fig.\ref{T1-U}) we can choose the initial 
condition for the domain wall assuming that (\ref{u1}) describes a
dS-Schwarzschild bulk with event and cosmological horizons when $M>0$
and $V_0>0$. 

We thus choose the initial condition for the domain wall inside this
region and where $r$ is a space coordinate. From
Fig.\ref{T1-U} let us notice that there are two small 
regions, $r_H \leq r < 0.593$ and $2.93 \leq r < r_C $, where
(\ref{condition1}) holds. The results are shown in  
Fig.\ref{T1-braneb}. We see that for region I the geodesics follow
the domain wall for a while and then decouple falling into the event
horizon. For region II all the geodesics and the
domain wall converge to the cosmological horizon $r_C$ independently
of the value of $\hat V_0$.

\subsubsection{$\hat \Lambda<0$, $M>0$}
This case describes an AdS-Schwarzschild bulk. The condition 
(\ref{condition1}) is fullfilled inside a very small range as we can 
see in Fig.\ref{T1-U3}(a). However,
all the geodesics fall into the event horizon after following some path
on the brane (see Fig.\ref{T1-U3}(b)).

\subsubsection{$\hat \Lambda>0$, $M<0$}
From Fig.\ref{T1-U2}(a) we choose the initial condition for the domain
wall equation of motion inside the region where (\ref{condition1})
holds. As we can see from Fig.\ref{T1-U2}(b), the domain wall and the
geodesic converge to the cosmological horizon $r_C$. However, after
some threshold initial velocity the geodesics diverge to the timelike
naked singularity. 
\subsubsection{$\hat \Lambda<0$, $M<0$}
In this case (\ref{dwmotion}) can only have solution when $k=-1$. This
is a topological black hole in an asymtotically AdS space.
From Fig.\ref{T1-U4} we see that there is no solution fulfilling  
(\ref{condition1}) between event and cosmological horizons. 
\subsection{Type II Solutions}

\begin{figure*}[htb!]
\begin{center}
\leavevmode
\begin{eqnarray}
\epsfxsize= 5.5truecm\rotatebox{-90}
{\epsfbox{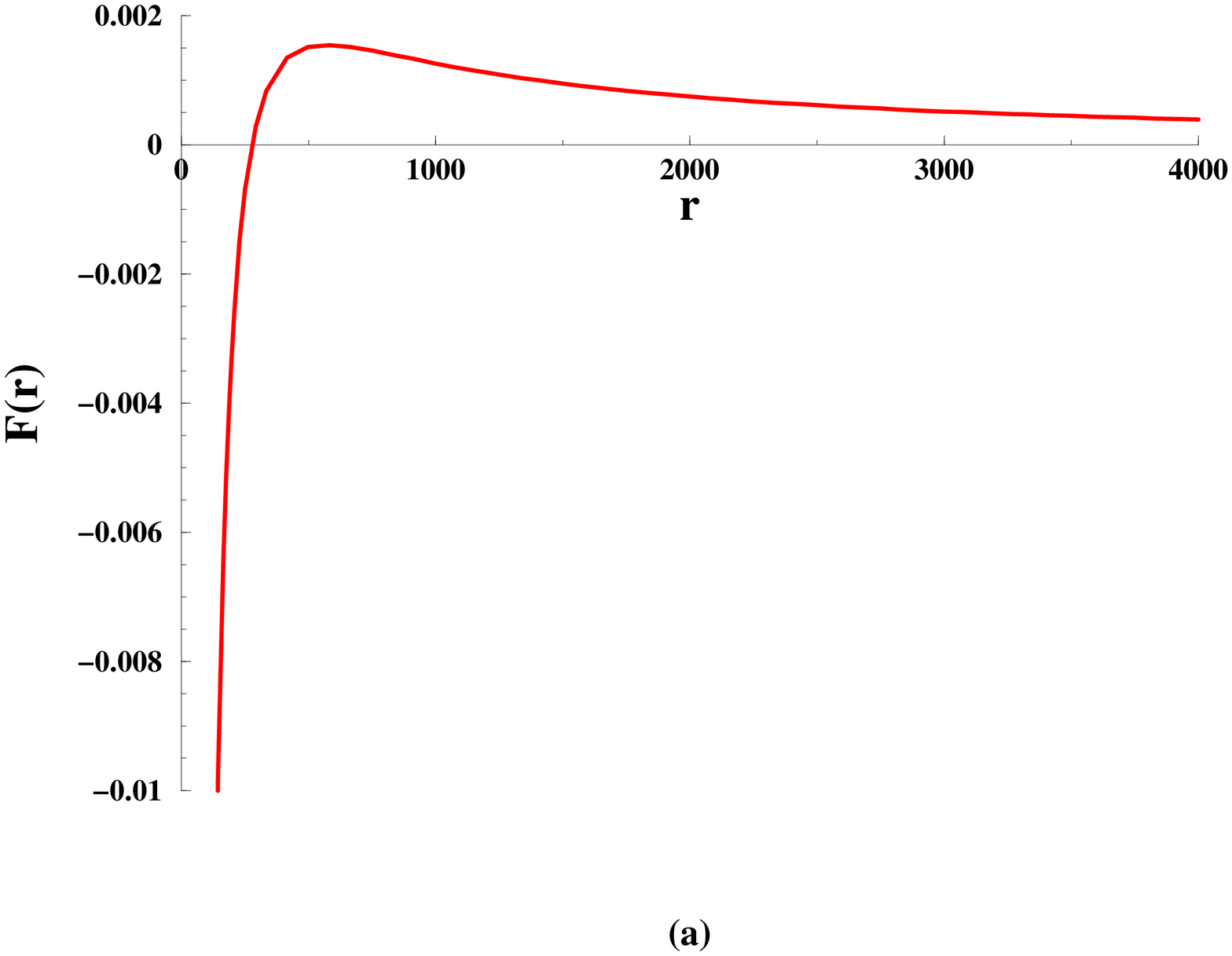}}\nonumber
\epsfxsize= 5.5truecm\rotatebox{-90}
{\epsfbox{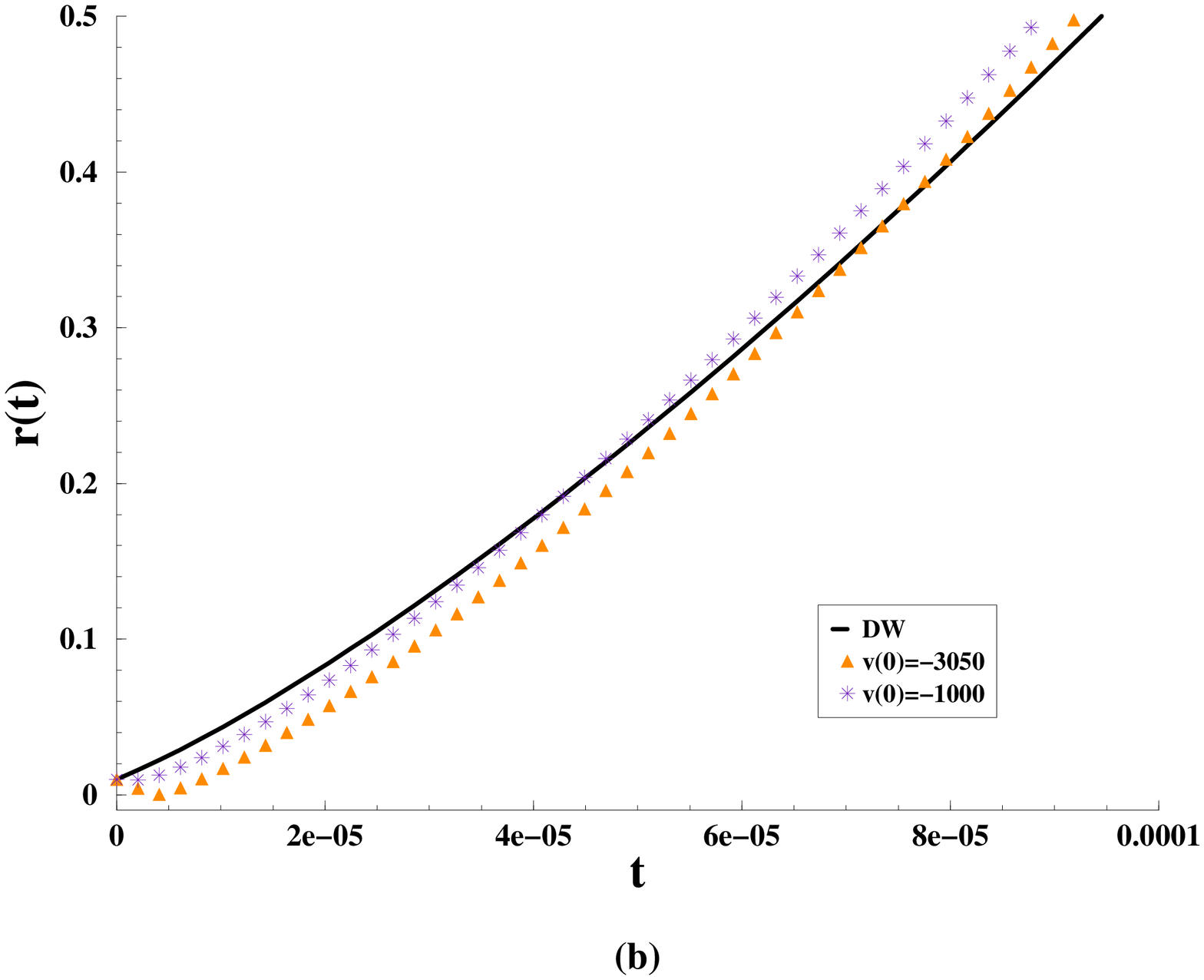}}\nonumber
\end{eqnarray}
\caption{(a) $F(r)$ with $\hat \Lambda >0$ and $M<0$ for type II 
solutions, (b) Domain wall motion  and geodesics for $V_0=1$,
$\hat V_0 =6$, $M=-10$ and $\beta=\sqrt{10}$.}
\label{T2-br1}
\end{center}
\end{figure*}
\begin{figure*}[htb!]
\begin{center}
\leavevmode
\begin{eqnarray}
\epsfxsize= 5.5truecm\rotatebox{-90}
{\epsfbox{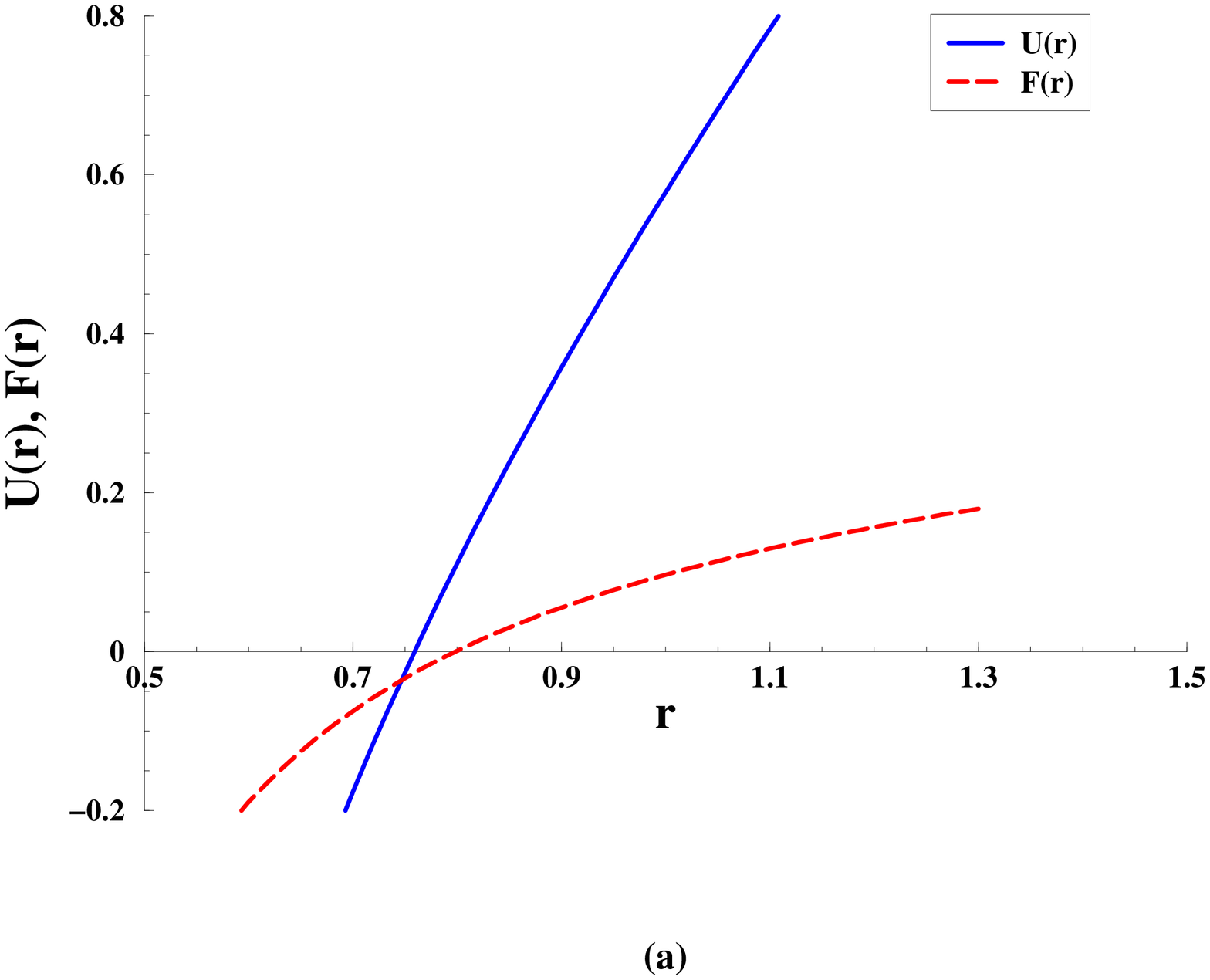}}\nonumber
\epsfxsize= 5.5truecm\rotatebox{-90}
{\epsfbox{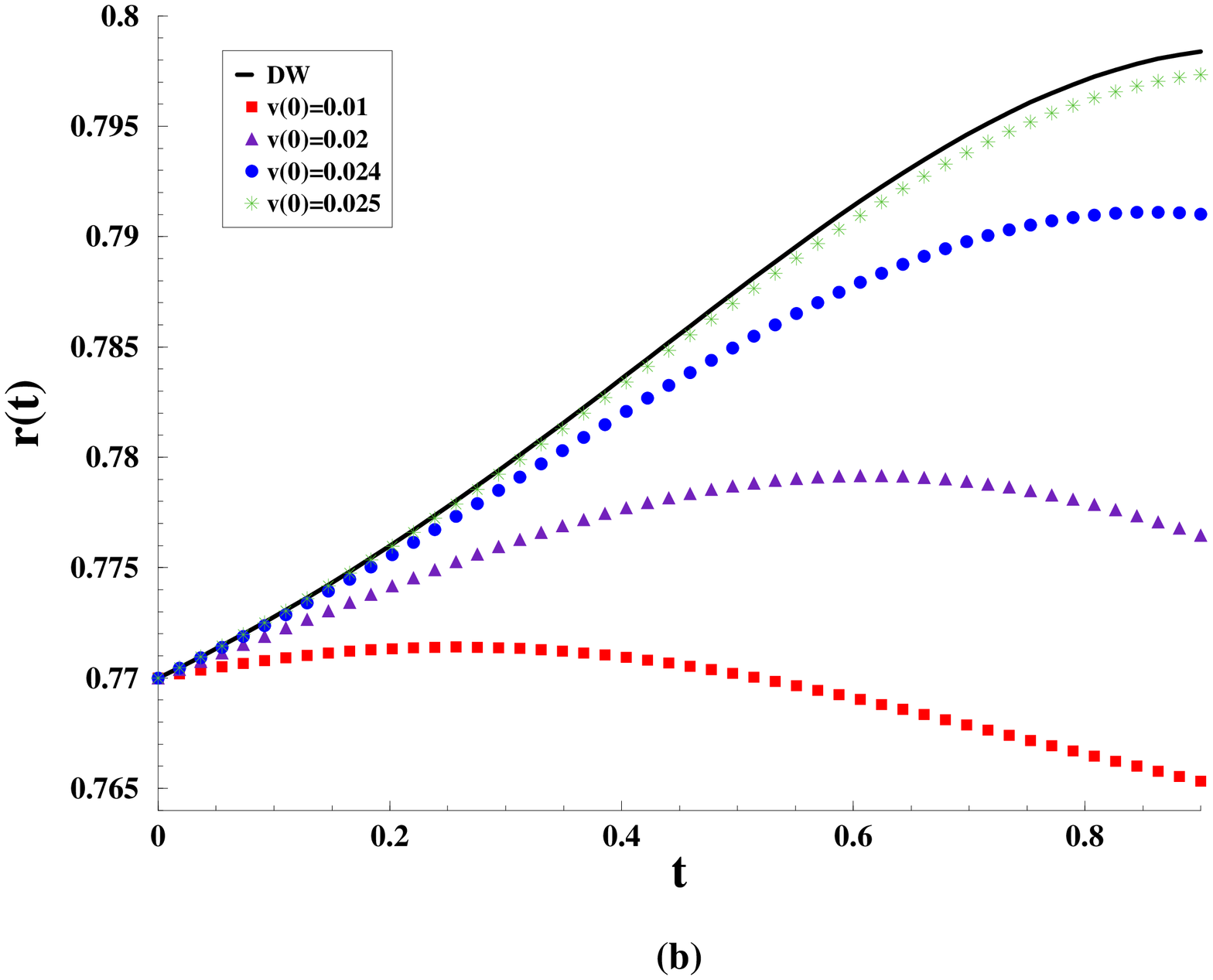}}\nonumber
\end{eqnarray}
\caption{(a) $U(r)$ and $F(r)$ with $\hat \Lambda <0$ and $M>0$ for type II 
solutions, (b) Domain wall motion  and geodesics for $V_0=-1$,
$\hat V_0 =1$, $M=1/10$ and $\beta=1/\sqrt{2}$.}
\label{T2-br2}
\end{center}
\end{figure*}
\begin{figure*}[htb!]
\begin{center}
\leavevmode
\begin{eqnarray}
\epsfxsize= 5.5truecm\rotatebox{-90}
{\epsfbox{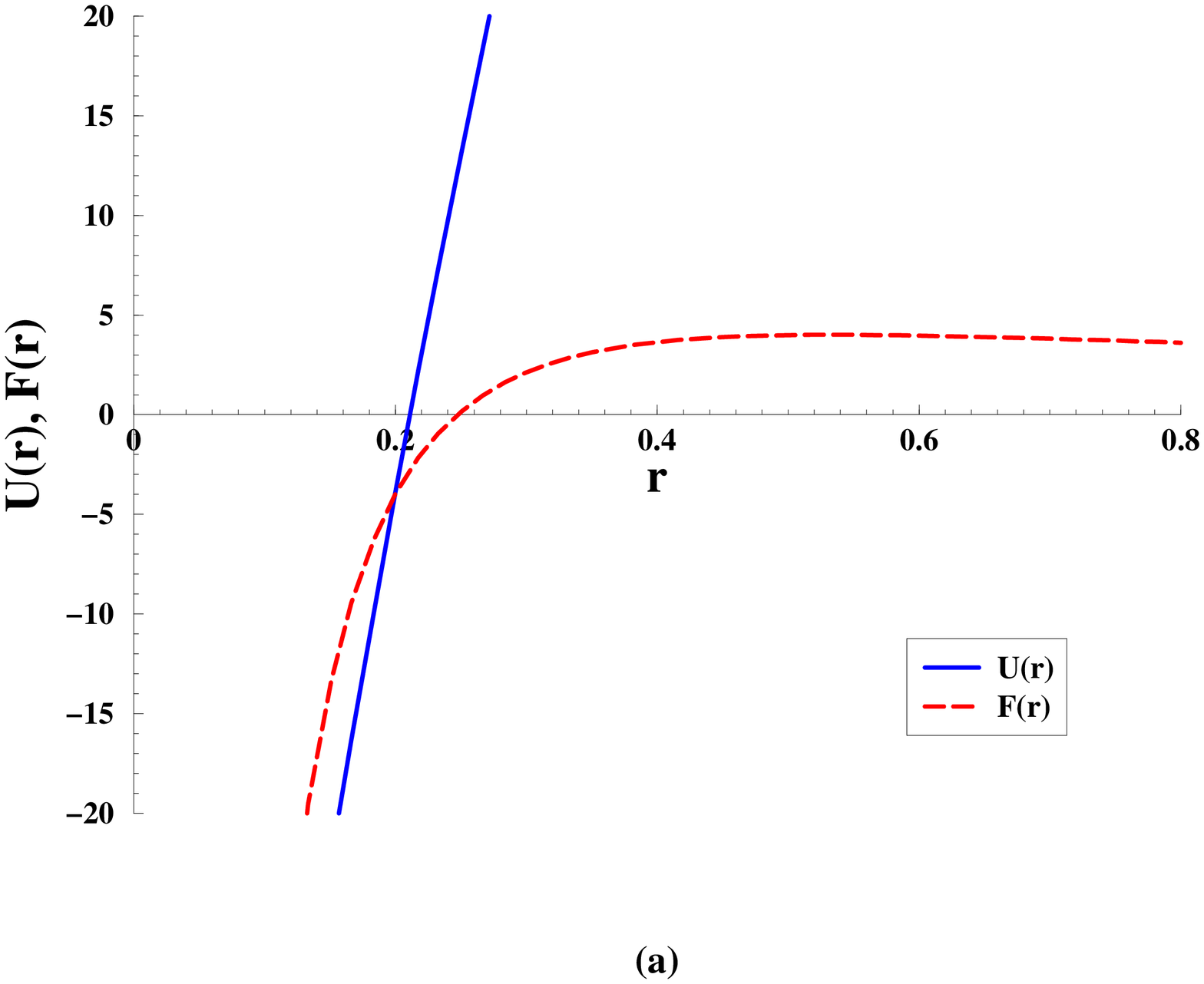}}\nonumber
\epsfxsize= 5.5truecm\rotatebox{-90}
{\epsfbox{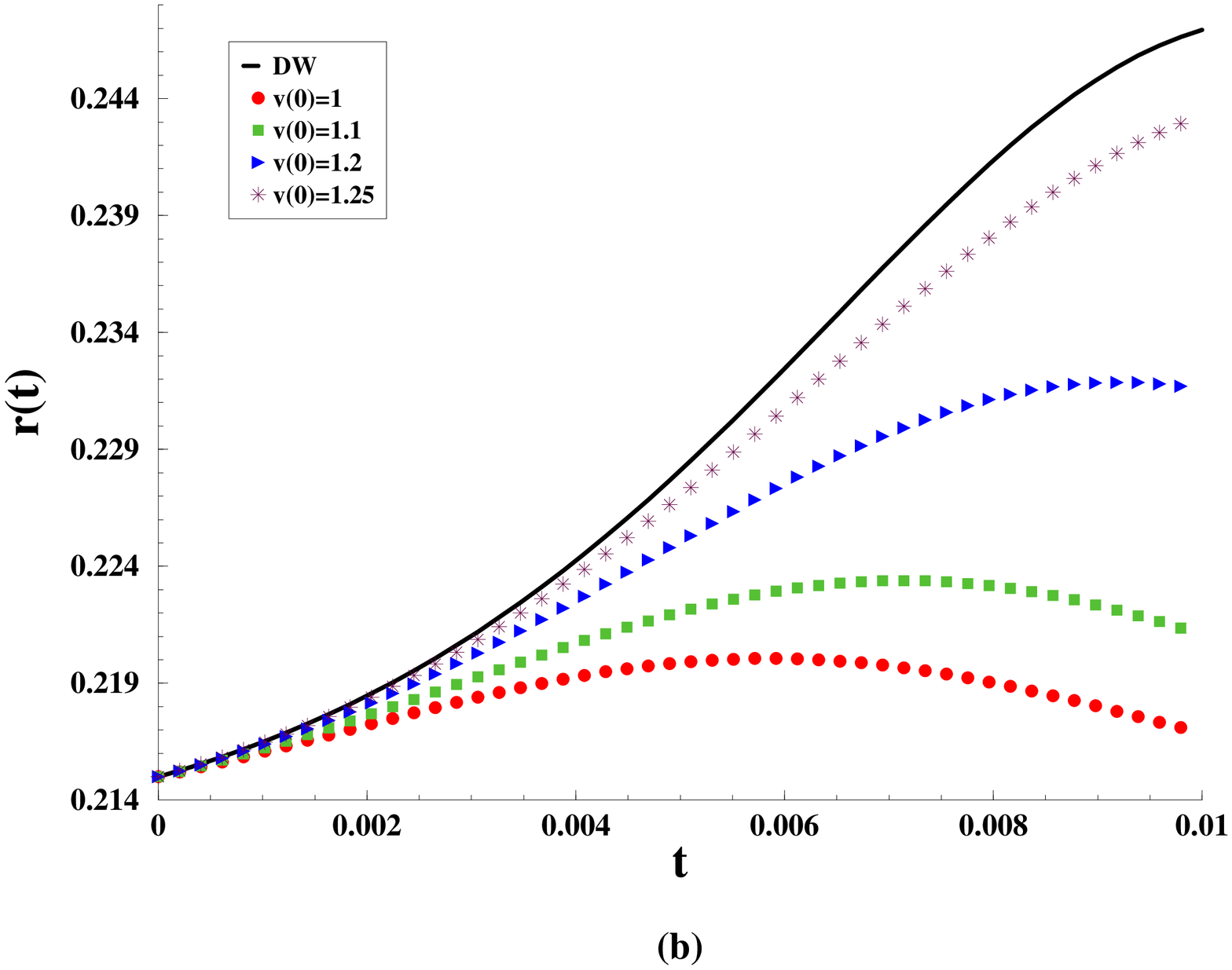}}\nonumber
\end{eqnarray}
\caption{(a) $U(r)$ and $F(r)$ with $\hat \Lambda <0$ and $M>0$ for type II 
solutions, (b) Domain wall motion  and geodesics for $V_0=-1$,
$\hat V_0 =1$, $M=10$ and $\beta=2$.}
\label{T2-br3}
\end{center}
\end{figure*}

The type II solutions have $\alpha =\beta/2$ and $k=0$. The metric is
given by
\begin{equation}\label{u2}
U(r) = (1 +b^2)^2 r^{2 \over {1+b^2}} \left( -2Mr^{-{{D-1-b^2}\over
{1+b^2}}} - {{2\Lambda}\over {(D-1-b^2)}} \right) \, ,
\end{equation}
and the scale factor is
\begin{equation}\label{R2}
R(r) = r^{1 \over {1+b^2}} \, ,
\end{equation}
where
\begin{eqnarray}
\Lambda &=& {{V_0 e ^{2b \phi_0}} \over {D-2}} \, , \label{lambda2} \\
b &=& {1 \over 2} \beta \sqrt{D-2} \, . \label{b2}
\end{eqnarray}

The potential is given by the expression
\begin{equation}\label{F2}
F(R) = -R^{2(1-b^2)} \left( M R^{-(D-1-b^2)} + \hat \Lambda \right) \; ,
\end{equation}
where 
\begin{equation}\label{ls2}
\hat\Lambda = {e^{2b\phi_0} \over {D-2}} \left( {V_0 \over {D-1-b^2}}
+ {{\hat V_0 ^2}\over {8(D-2)}} \right) \; .
\end{equation}

There are twelve cases from which we choose those ones where $r$ is a
spatial coordinate. When $b^2<D-1$, $r$ is a spatial coordinate if
$V_0<0$. When $b^2>D-1$, $r$ is spatial if $M<0$.

We should also rewrite (\ref{condition1}) as
\begin{equation}\label{condition2}
F(r) \leq 0 \quad \hbox{and} \quad U(r)>0 \, . 
\end{equation}

\subsubsection{$\hat\Lambda>0$, $M<0$, $b^2>D-1$}

In this case $U(r)$ is always positive, whereas $F(r)$ is negative for
small $r$. From Fig.\ref{T2-br1} we see that some microscopic shortcuts
appear in the very beginning of the 
solution and after crossing the domain wall they escape to infinity.

\subsubsection{$\hat\Lambda<0$, $M>0$, $b^2<1$}

This case describes a black $(D-2)$ brane solution in AdS space. Here there is
a very small region where (\ref{condition2}) holds after 
the event horizon as we can see from Fig.\ref{T2-br2}. We show the
entire domain wall solution and we see that geodesics follow it
and then decouple to fall into the event horizon at later times.

\subsubsection{$\hat\Lambda<0$, $M>0$, $1<b^2<D-1$}

This case is also a black brane in AdS space. The region where
(\ref{condition2}) is respected is shown in Fig.\ref{T2-br3}. As in
the previous case all the geodesics follow the domain wall and at
later times fall into the event horizon.

\subsubsection{$\hat\Lambda<0$, $M<0$}

As $F(r)$ is always positive for all $b^2$, no solutions to
(\ref{eq3}) exist. 

\subsection{Type III Solutions} 

\begin{figure*}[htb!]
\begin{center}
\leavevmode
\begin{eqnarray}
\epsfxsize= 5.truecm\rotatebox{-90}
{\epsfbox{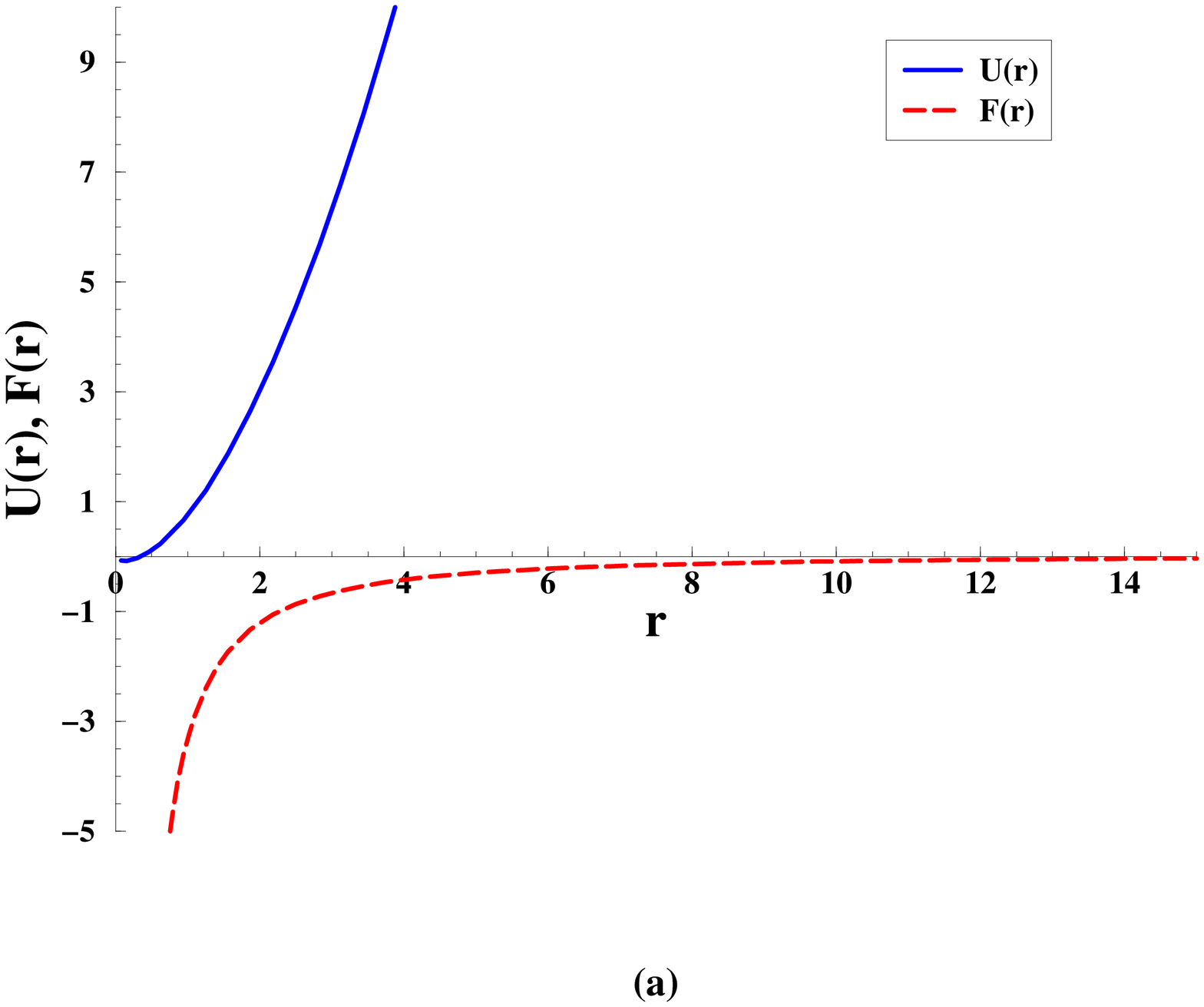}}\nonumber
\epsfxsize= 5.truecm\rotatebox{-90}
{\epsfbox{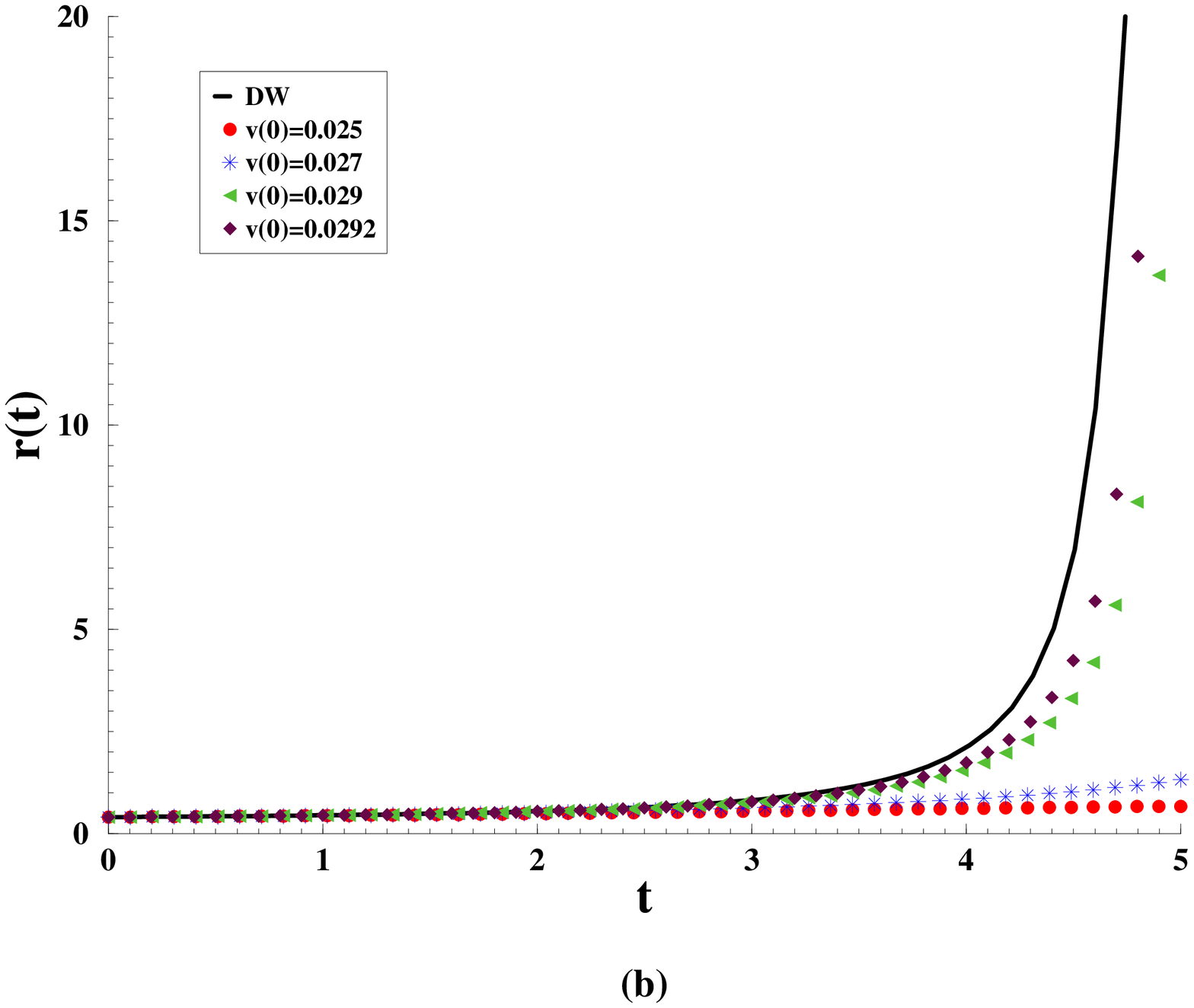}}\nonumber
\end{eqnarray}
\caption{(a) $U(r)$ and $F(r)$ for Type III solutions with $k=-1$, $M=1/10$ and
$\beta^2<{1 \over {(D-2)}}$, (b) Domain Wall motion  and
geodesics for $V_0=-1$, $\hat V_0=1$, $\phi_0=1$ and
$\beta=1/\sqrt{6}$.}
\label{T3-branea}

\begin{eqnarray}
\epsfxsize= 5.truecm\rotatebox{-90}
{\epsfbox{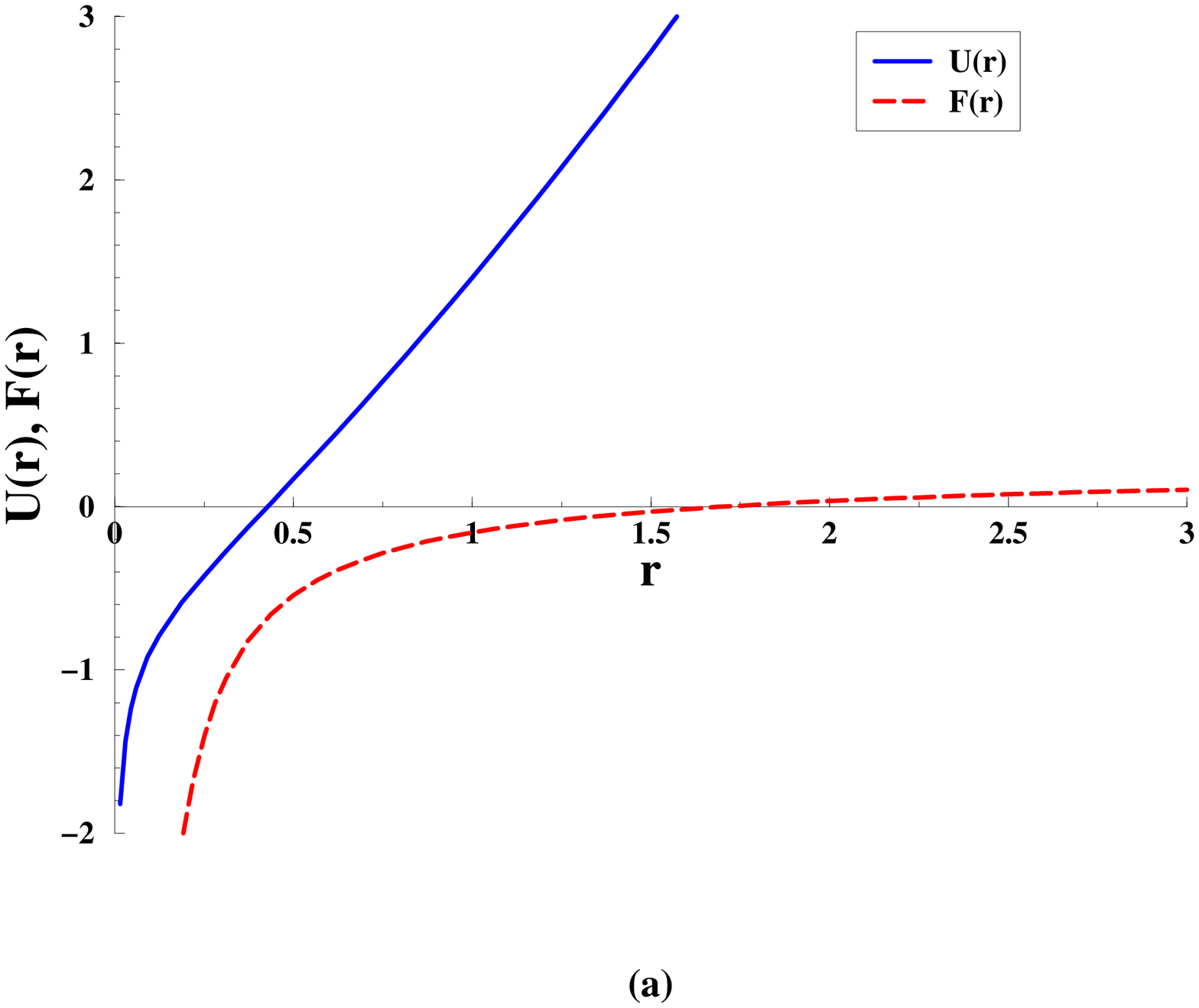}}\nonumber
\epsfxsize= 5.truecm\rotatebox{-90}
{\epsfbox{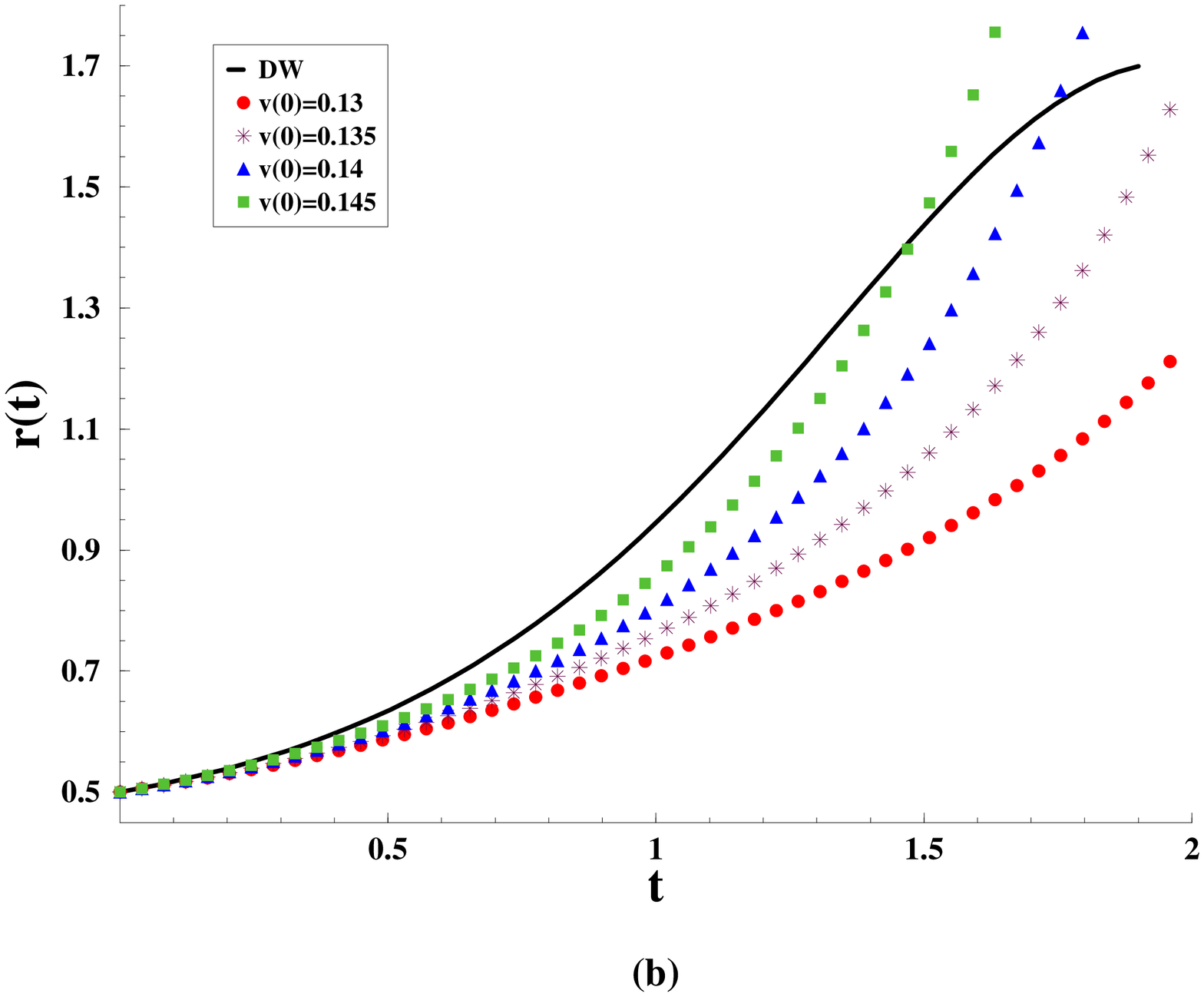}}\nonumber
\end{eqnarray}
\caption{(a) $U(r)$  and $F(r)$  for $k=-1$, $M=1/10$ and
$V_0<0$ in Type III solutions, 
(b) Domain wall motion  and geodesics with $V_0=-1$,
$\hat V_0=1$, $\phi_0=1$ and $\beta=1/\sqrt{2}$.}
\label{T3-braneb}
\end{center}
\end{figure*}
\begin{figure*}[htb!]
\begin{center}
\leavevmode
\begin{eqnarray}
\epsfxsize= 5.5truecm\rotatebox{-90}
{\epsfbox{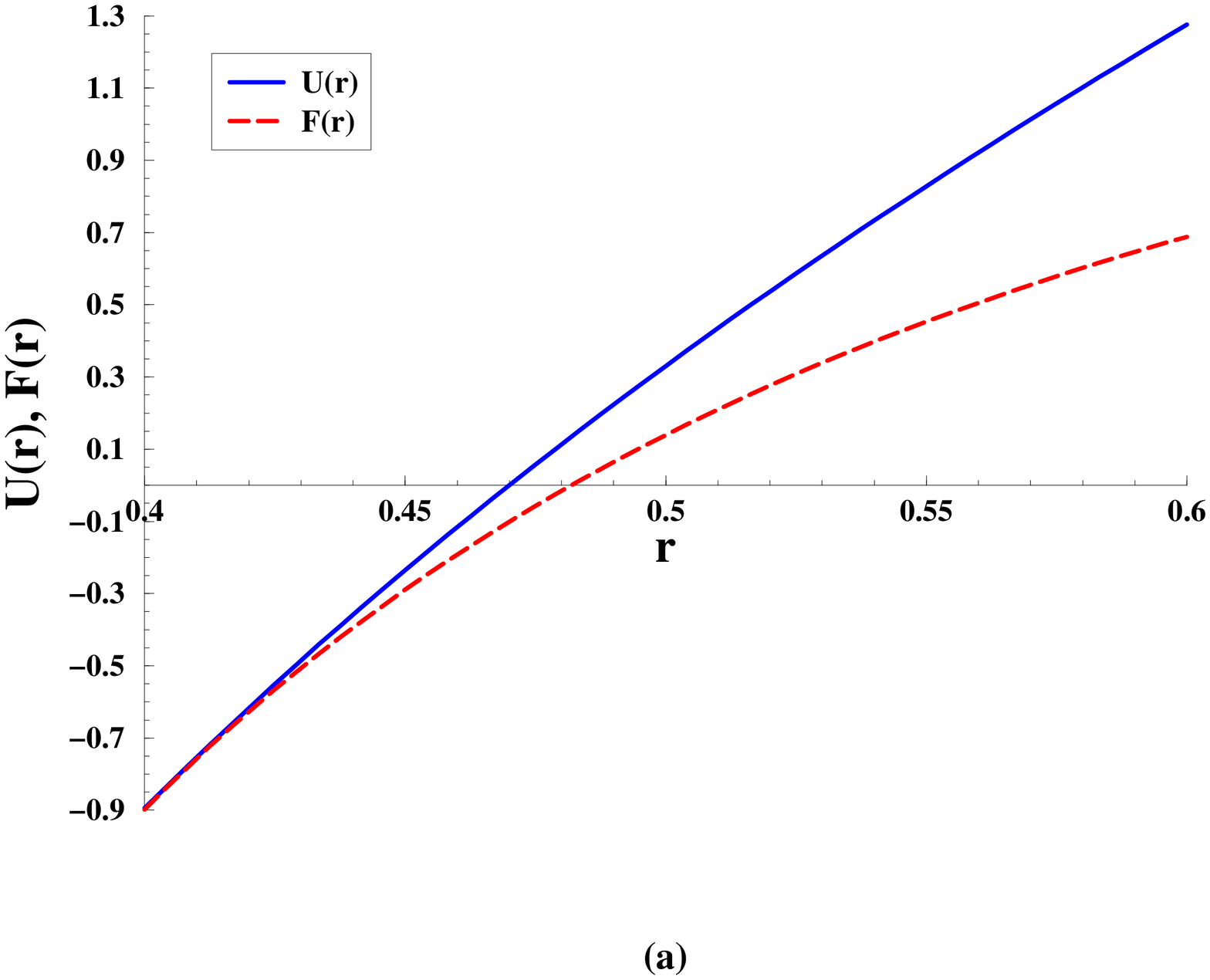}}\nonumber
\epsfxsize= 5.5truecm\rotatebox{-90}
{\epsfbox{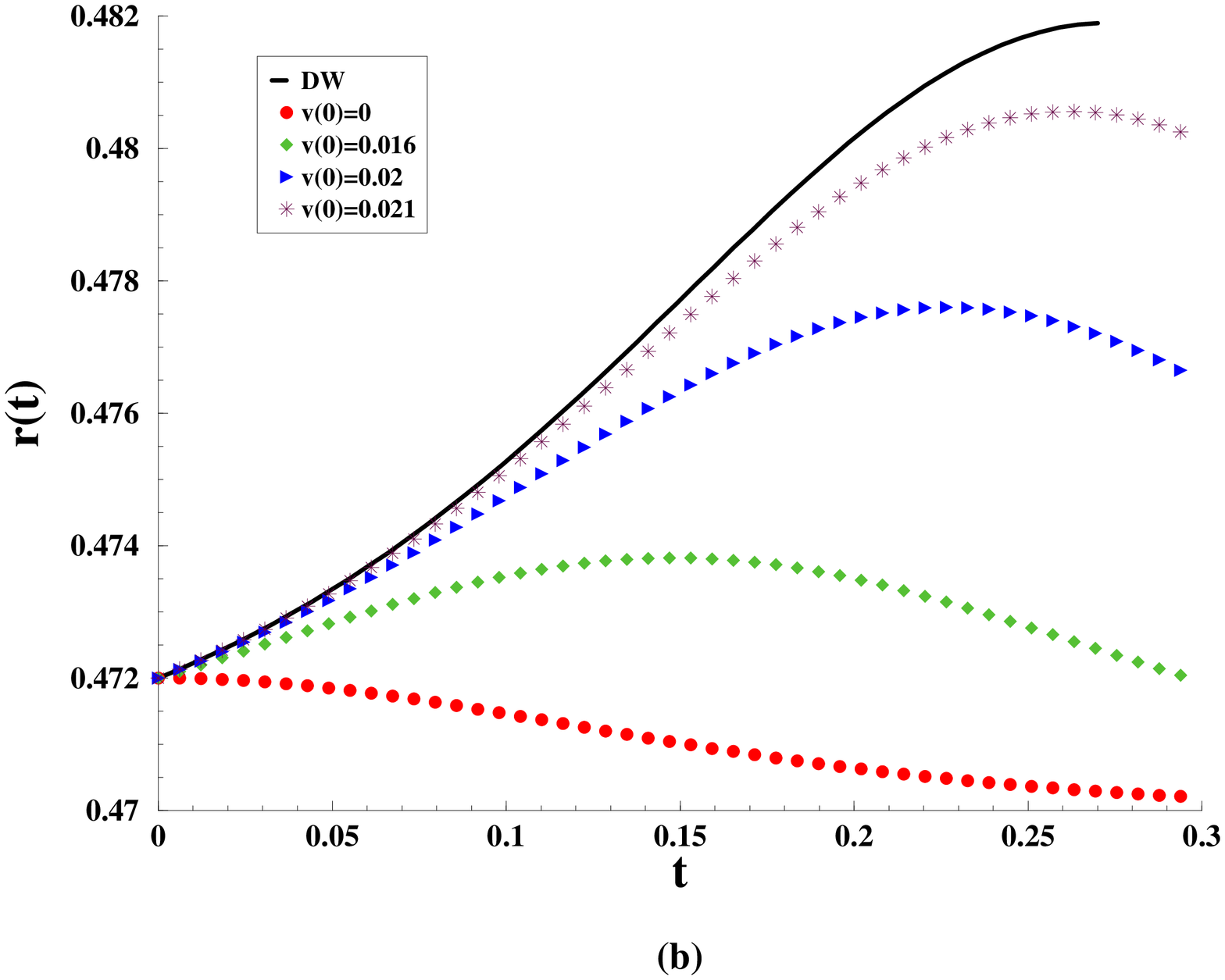}}\nonumber\\
\epsfxsize= 5.5truecm\rotatebox{-90}
{\epsfbox{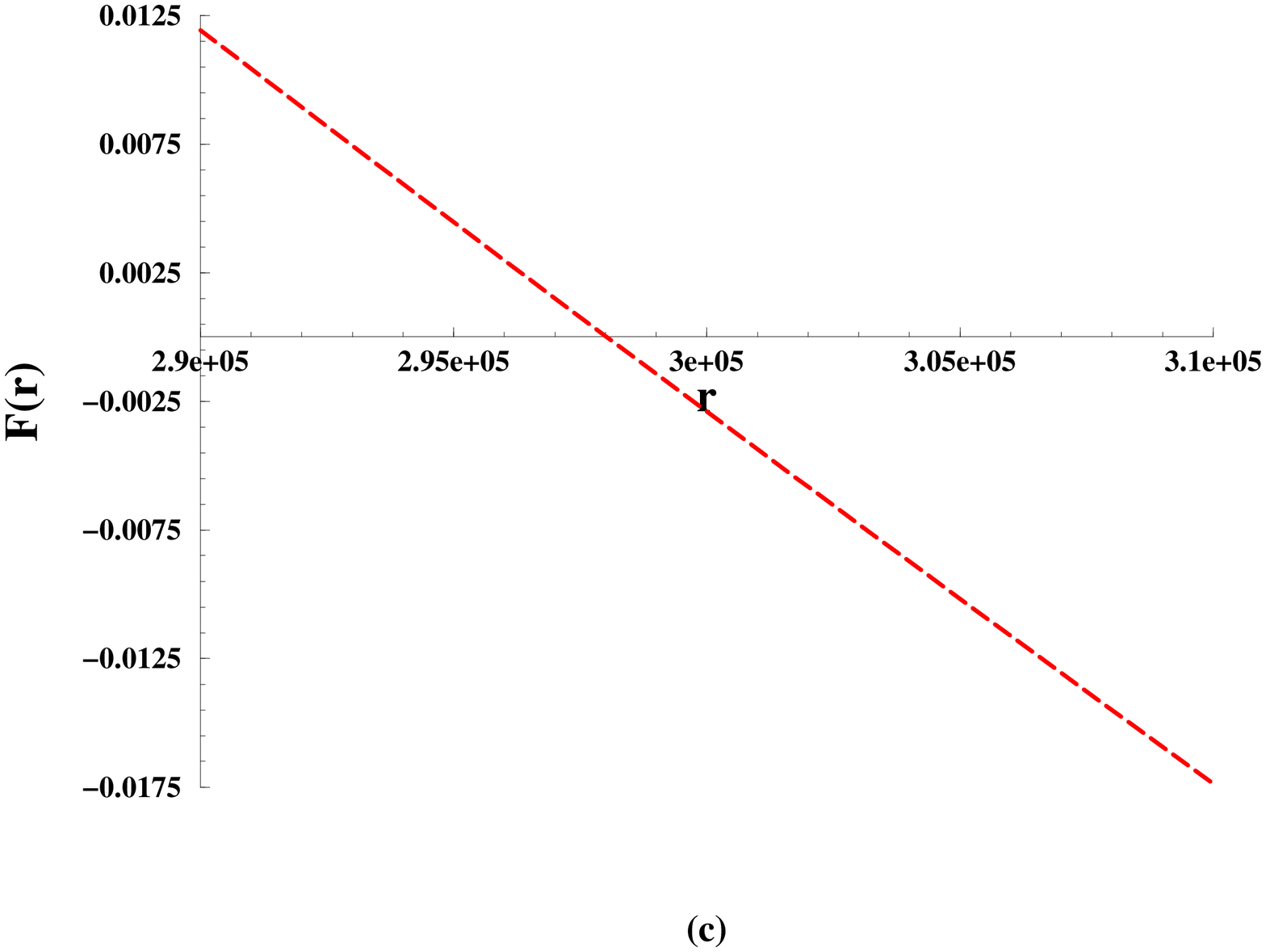}}\nonumber
\epsfxsize= 5.5truecm\rotatebox{-90}
{\epsfbox{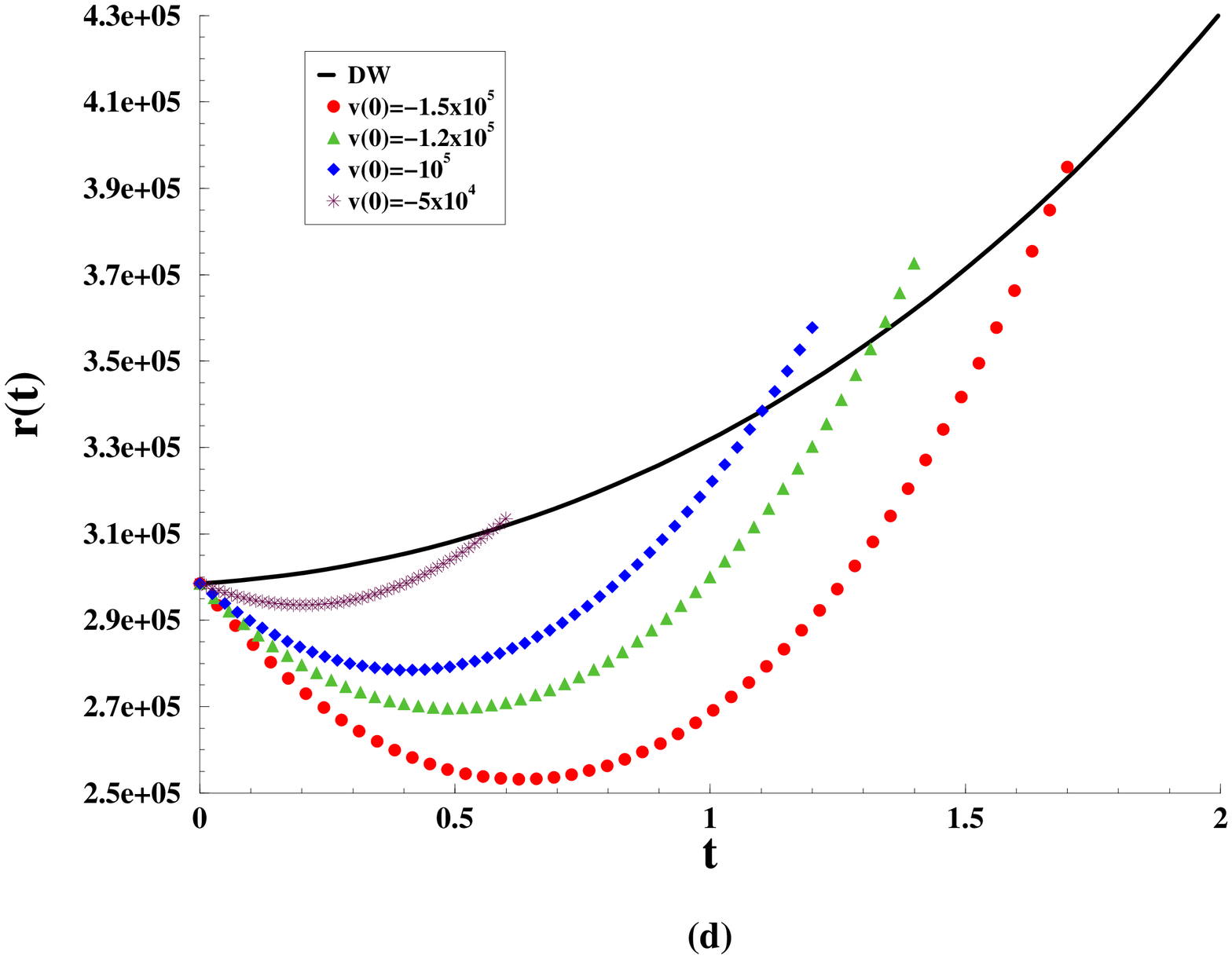}}\nonumber
\end{eqnarray}
\caption{(a) $U(r)$  and $F(r)$  for $M=1/10$ and $V_0<0$
in Type III solutions, 
(b) Domain wall motion  and geodesics with $V_0=-1$,
$\hat V_0=1$, $\phi_0=1$ and $\beta=\sqrt{5}/2$, (c) $F(r)$ in the
region of interest, (d) Domain wall motion  and geodesics under
the same conditions as (b).}
\label{T3-braned}
\end{center}
\end{figure*}

\begin{figure*}[htb!]
\begin{center}
\leavevmode
\begin{eqnarray}
\epsfxsize= 5.5truecm\rotatebox{-90}
{\epsfbox{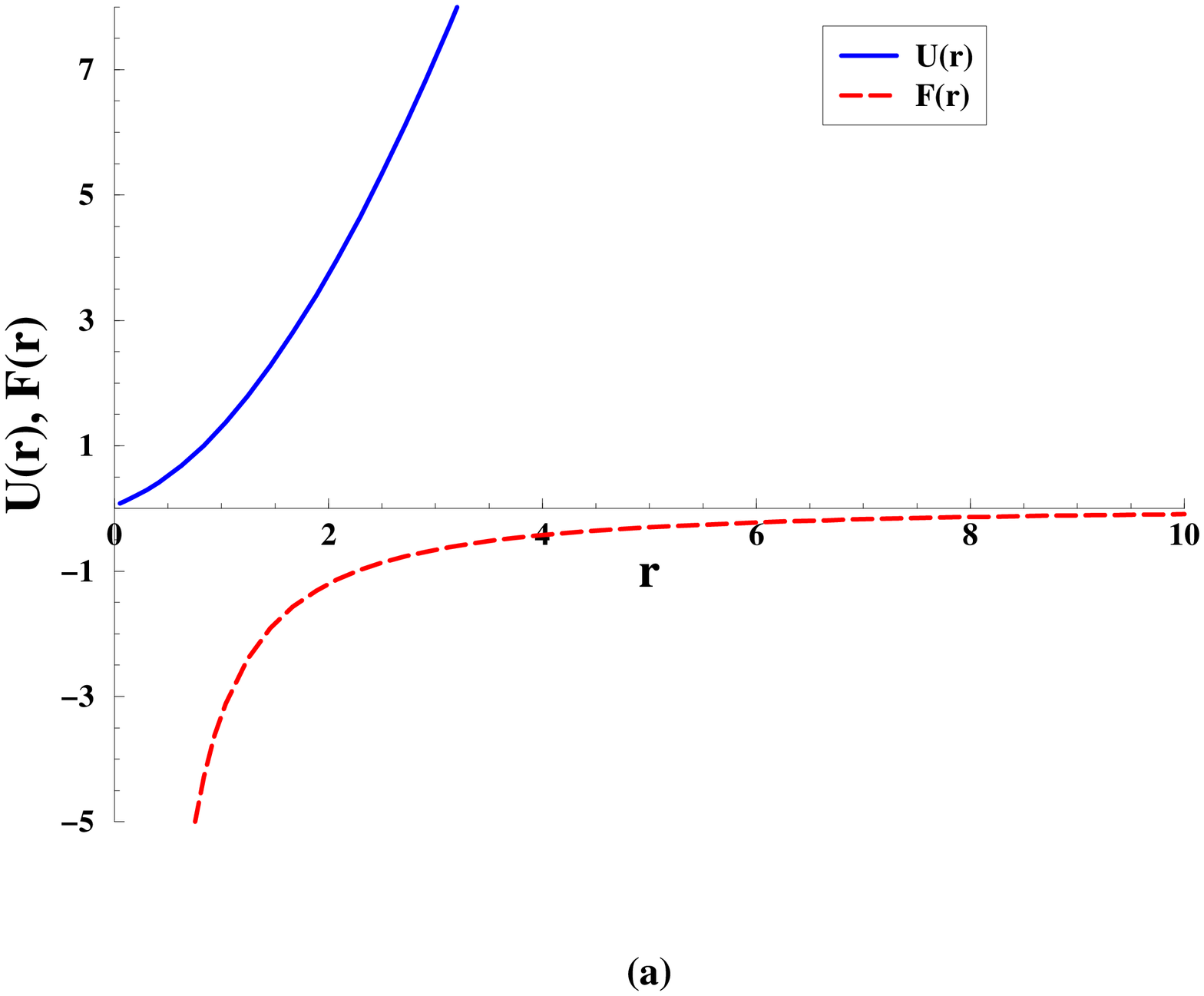}}\nonumber
\epsfxsize= 5.5truecm\rotatebox{-90}
{\epsfbox{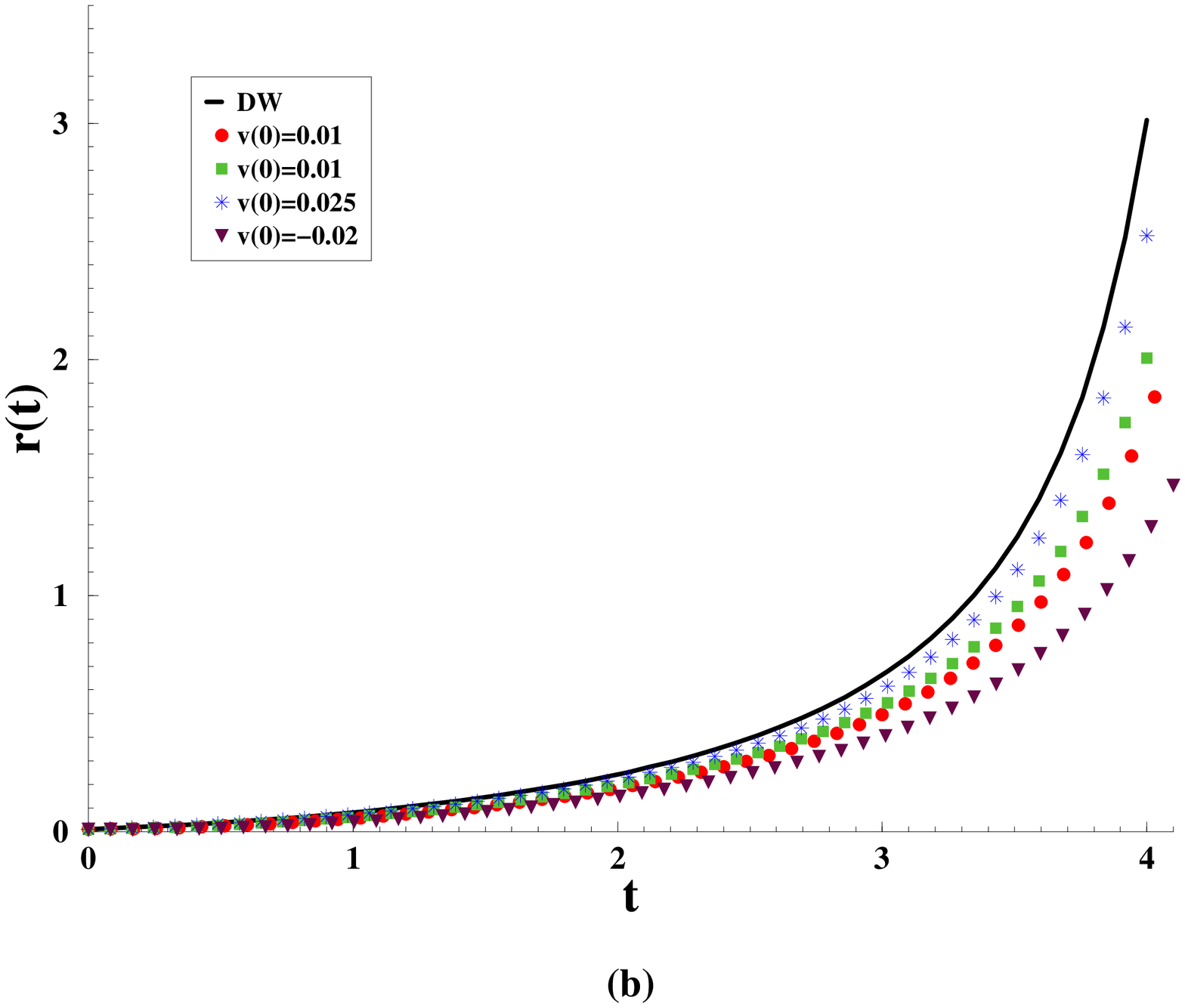}}\nonumber
\end{eqnarray}
\caption{(a) $U(r)$ and $F(r)$ for type III solutions when $k=-1$, $M=-1/10$
and $b^2<{1 \over {(D-2)}}$, (b)
Domain wall motion  and geodesics for $V_0=-1$, $\hat V_0=1$,
$\phi_0=1$ and $\beta=1/\sqrt{6}$.}
\label{T3-branee}
\end{center}
\end{figure*}
\begin{figure*}[htb!]
\begin{center}
\leavevmode
\begin{eqnarray}
\epsfxsize= 5.5truecm\rotatebox{-90}
{\epsfbox{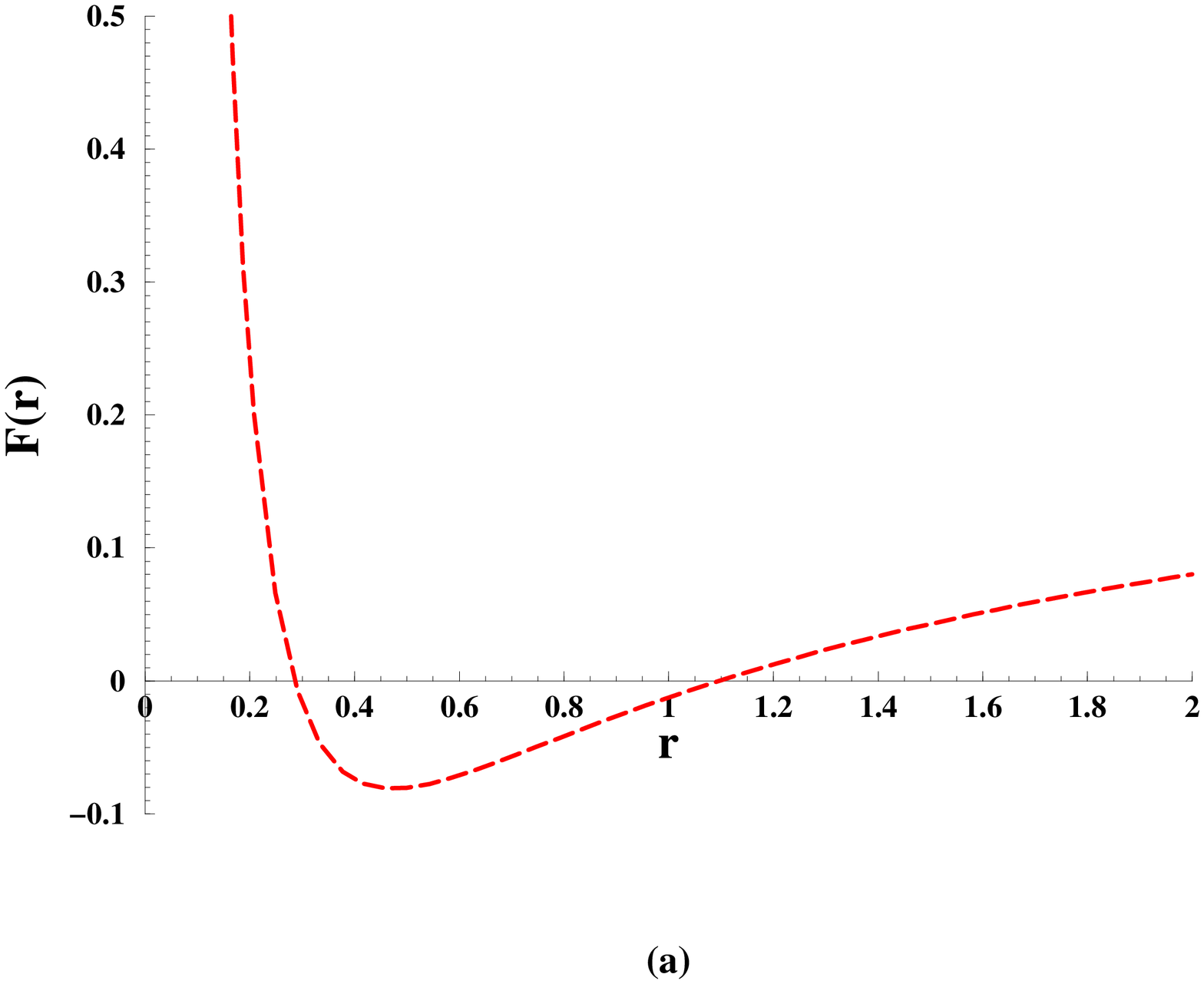}}\nonumber
\epsfxsize= 5.5truecm\rotatebox{-90}
{\epsfbox{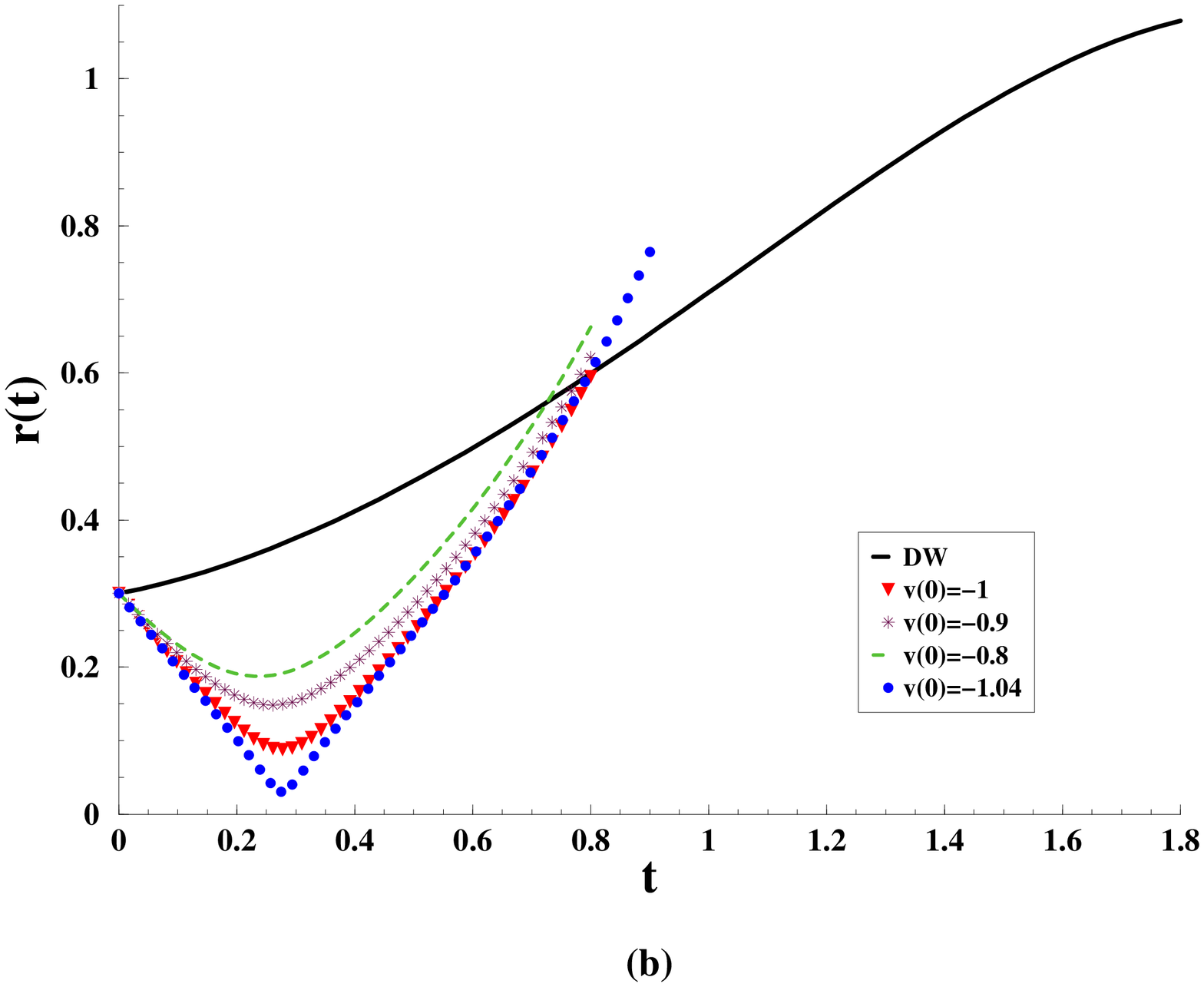}}\nonumber
\end{eqnarray}
\caption{(a) $F(r)$ for type III solutions when $k=-1$, $M=-1/10$ and ${1 \over
{(D-2)}}<b^2<1$, (b)
Domain wall motion  and geodesics for $V_0=-1$, $\hat V_0=1$,
$\phi_0=1$ and $\beta=1/\sqrt{2}$.}
\label{T3-branef}
\end{center}
\end{figure*}
\begin{figure*}[htb!]
\begin{center}
\leavevmode
\begin{eqnarray}
\epsfxsize= 5.5truecm\rotatebox{-90}
{\epsfbox{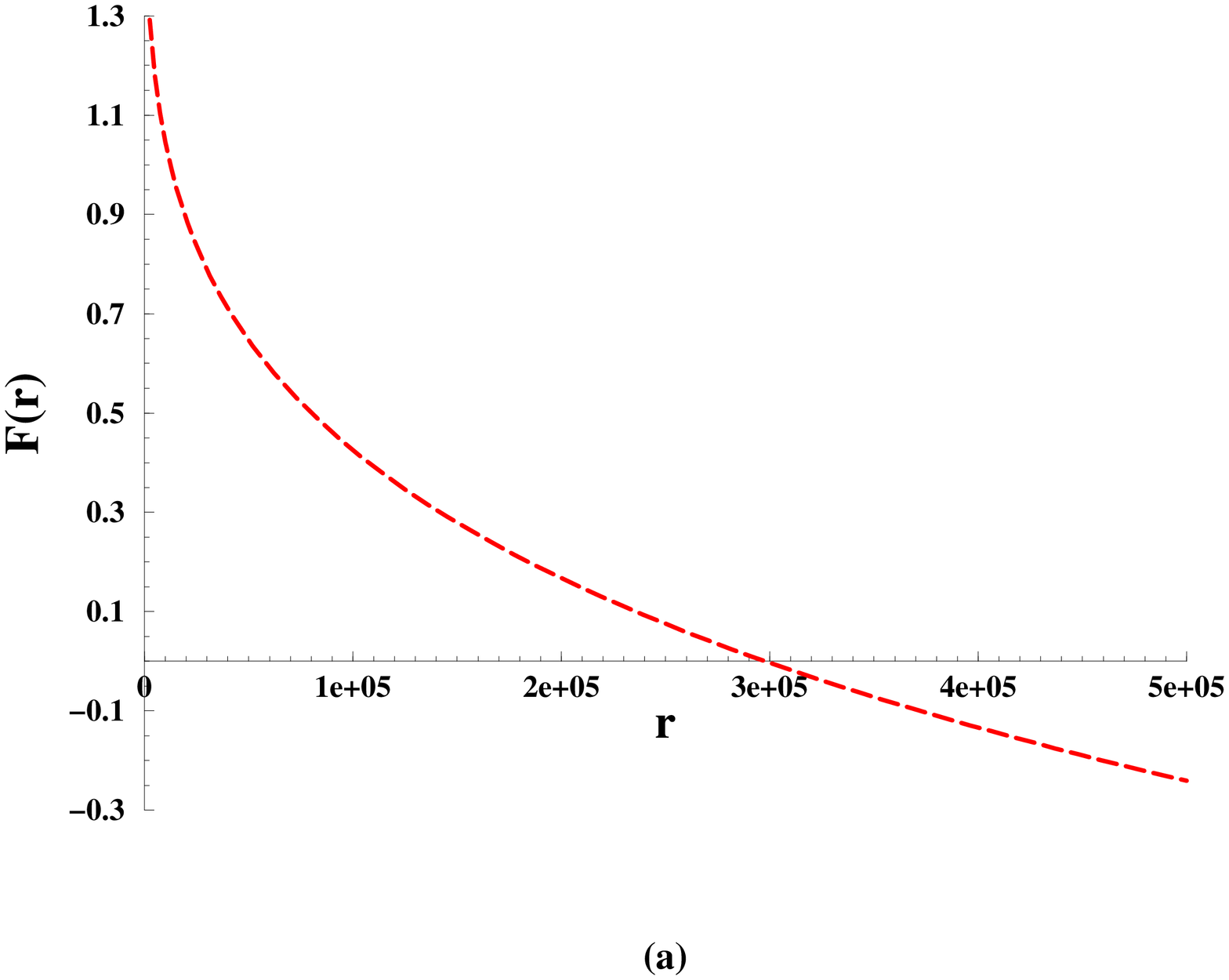}}\nonumber
\epsfxsize= 5.5truecm\rotatebox{-90}
{\epsfbox{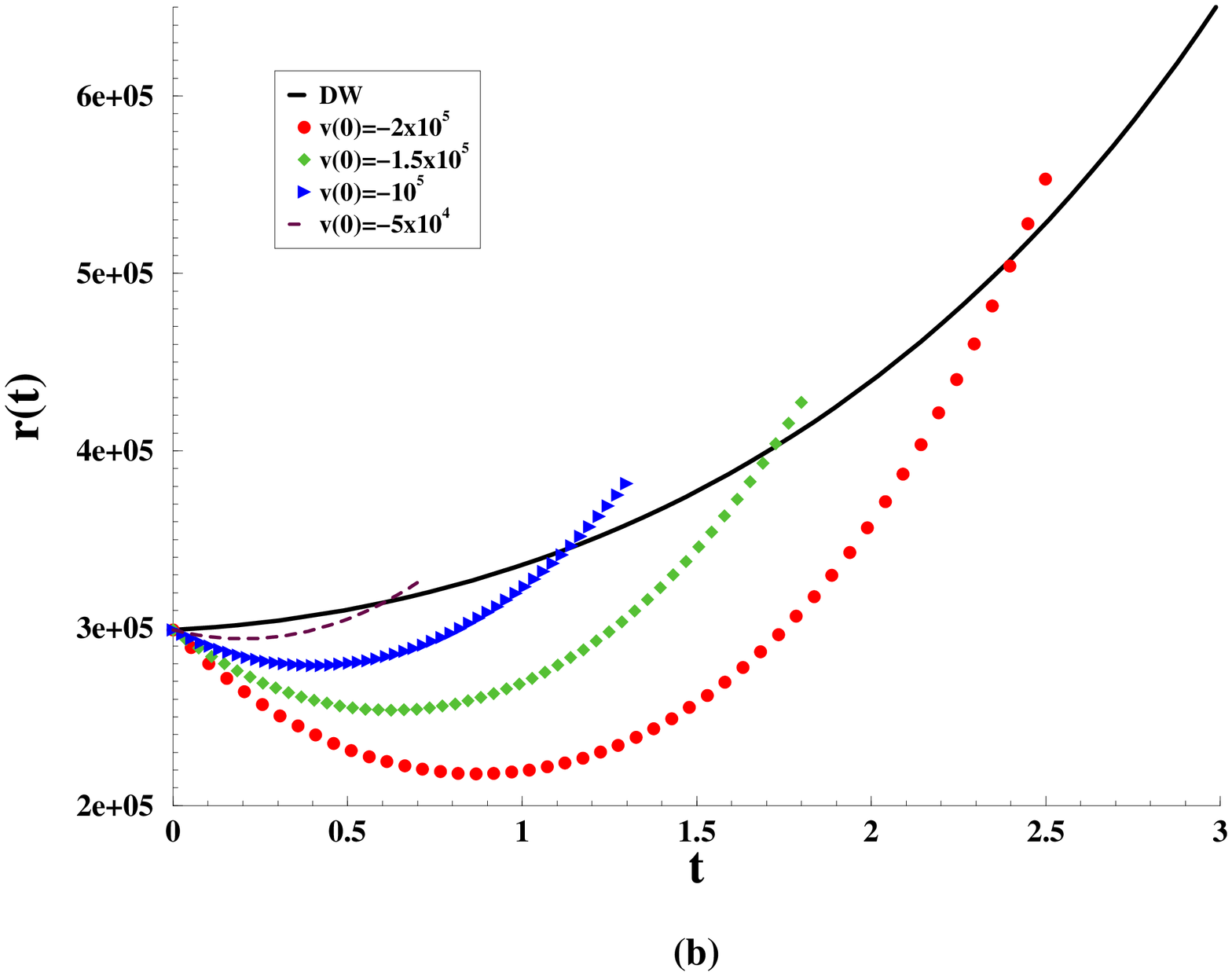}}\nonumber
\end{eqnarray}
\caption{(a) $F(r)$ for type III solutions when $M=-1/10$ and $b^2>1$, (b)
Domain wall motion  and geodesics for $V_0=-1$, $\hat V_0=1$,
$\phi_0=1$ and $\beta=\sqrt{5}/2$.}
\label{T3-braneg}
\end{center}
\end{figure*}

The type III solutions have $\alpha = {2\over {\beta(D-2)}}$.
In this case, the metric is given by
\begin{equation}\label{u3}
U(r) = (1+b^2)^2 r^{2\over {1+b^2}} \left( -2Mr^{-{1+b^2(D-3)}\over
    {1+b^2}} - {{2\Lambda} \over {(1+b^2(D-3))}} \right) \, ,
\end{equation}
and the scale factor is 
\begin{equation}\label{R3}
R(r) = \gamma r^{b^2 \over {1+b^2}} \, ,
\end{equation}
where
\begin{equation}\label{gamma3}
\gamma = \left( {{(D-3)}\over{2k\Lambda(1-b^2)}} \right)^{1\over 2}
\, .
\end{equation}
The values of $\Lambda$ and $b$ are the same as those given in 
(\ref{lambda2}) and (\ref{b2}).

The potential $F(R)$ is
\begin{eqnarray}
F(R) &=& - {{(D-3)b^4} \over {2k(1-b^2)(1+b^2(D-3))}} - M \gamma^2 b^4
\left({R \over \gamma} \right) ^{-\left(D-3+{1\over b^2}\right)} -
\nonumber \\
&&- {{\hat V_0 ^2 e ^{{2\phi_0}\over b} \gamma^2} \over {8(D-2)^2}}
\left({R \over \gamma} \right) ^{-2\left({1\over b^2} -1\right)} \,
.\label{f3}
\end{eqnarray}

If $V_0>0$, $r$ turns out to be a time coordinate, while for $V_0<0$, it 
is a spatial coordinate. From the twelve cases shown in \cite{chre} we 
choose those where it is a spatial coordinate. 
For all these solutions the condition (\ref{condition2}) applies.

\subsubsection{$V_0<0$, $M>0$, $b^2<{1 \over {(D-1)}}$}
This case describes a topological black hole in AdS space. From
Fig.\ref{T3-branea} we can see the region where (\ref{condition2})
holds.  
There are no shortcuts in this interval, and all the geodesics follow the 
brane and then either diverge to infinity or fall into the event horizon.

\subsubsection{$V_0<0$, $M>0$, ${1 \over {(D-1)}}<b^2<1$}
We again have a topological black hole in AdS space. There is a small
interval where (\ref{eq3}) has solution as we can see 
from Fig.\ref{T3-braneb}(a). Our results are shown in
Fig.\ref{T3-braneb}(b). Notice that the domain wall equation of
motion has a solution only inside the interval shown there. This means that
only a group of geodesics with initial velocity $\dot r(0)> v_c$ can
meet the domain wall after a roundabout in the bulk. It is also shown that 
negative initial velocities force geodesics to fall into the event
horizon.

\subsubsection{$V_0<0$, $M>0$, $b^2>1$}
The black hole in AdS space appearing here has round spatial
section. In this case $U(r)$ is always positive (then $r$ is always a
spatial coordinate); however, as we must fulfill
(\ref{condition2}), we should notice that $F(r) \leq 0$ for $r\geq
3*10^5$. We found that shortcuts
are possible for several initial velocities if $M=0$.

The case $M>0$ is shown in Fig.\ref{T3-braned}. We have two regions of
interest after the event horizon depending only on the sign of $F(r)$
since $U(r)$ is positive in this range. In the first region
all the geodesics initially follow the brane and fall into the event
horizon at later times. In the second region we have shortcuts again
for several initial velocities. 

\subsubsection{$V_0<0$, $M<0$, $b^2<{1 \over {(D-1)}}$}
Here $U(r)$ is always positive while $F(r)$ is negative in the range
shown in Fig.\ref{T3-branee}. The domain wall and the geodesics
diverge after some time near the end of the range where (\ref{eq3})
has a solution. 

\subsubsection{$V_0<0$, $M<0$, ${1 \over {(D-1)}}<b^2<1$}
In this case $U(r)$ is always positive while $F(r)$ is negative for a small 
range as seen in Fig.\ref{T3-branef}. There are several shortcuts in the 
region where the domain wall equation of motion has solution; nevertheless, 
there is a threshold velocity after which the geodesics can not
return. As we can notice from Fig.\ref{T3-branef}(b), the last
curve displayed here ($v(0)=-1.04$) can not be consider a real shortcut
since it is not a continuous solution of
the geodesic equation, but represents a transition between shortcuts
and geodesics falling into the naked singularity. 

\subsubsection{$V_0<0$, $M<0$, $b^2>1$}
Now $U(r)$ is always positive and $F(r)$ will determine the
initial condition for the domain wall equation of motion. As we can
see from Fig.\ref{T3-braneg}, several shortcuts appear.

\section{Domain Wall Time and Time Delays}

The time delay between the photon traveling on the domain wall and the
gravitons traveling in the bulk \cite{abdacasali} can be calculated as
follows. Since the signals cover the same distance, 
\begin{equation}\label{timedelay}
\int {{d\tau_\gamma}\over {r(\tau_\gamma)}} = \int {{dt_g}\over
{r_g(t_g)}} \sqrt{U(r_g) - {{\dot r_g(t)^2}\over {U(r_g)}}} \, ,
\end{equation}
the difference between photon and graviton time of flight can
approximately be written as
\begin{equation}
{{\Delta\tau}\over {r}} \simeq \int_0 ^{\tau_f+\Delta\tau}
{{d\tau_\gamma}\over {r(\tau_\gamma)}} - \int_0 ^{\tau_f}
{{d\tau_g}\over{r(\tau_g)}} \, , \nonumber
\end{equation}
or in terms of the bulk time
\begin{equation}\label{delay}
\Delta\tau \simeq r(t_f) \int_0 ^{t_f} dt \left({{1}\over {r_g(t)}}
\sqrt{U(r_g)- {{\dot r_g(t)^2}\over {U(r_g)}}} - {1 \over {r(t)}}
{{d\tau}\over{dt}} \right) \, .
\end{equation}
The elapsed bulk time $t$, 
the corresponding domain wall time $\tau$ and the delays (\ref{delay})
are shown in Table \ref{sc} for all the shortcut examples considered
in the present paper. Notice that $\Delta\tau <0$ for the last
geodesic solution in Fig.\ref{T3-branef} what, in fact, shows that this
curve can not be considered a shortcut (it is probably falling into
the naked singularity).

\begin{table}[t!]
\begin{center}
\begin{tabular}{|c|c|c|c|}\hline\hline
\multicolumn{4}{|c|}{{\bf Type II Shortcuts }}\\ \hline \hline
Conditions                     &$t$          &$\tau$
&$\Delta\tau$\\ \hline 
Fig.\ref{T2-br1}, $v(0)=-3050$ &$7.5\times 10^{-5}$&$0.0025$&$0.0466$
\\ \hline  
Fig.\ref{T2-br1}, $v(0)=-1000$ &$4.4\times 10^{-5}$&$0.0014$&$0.0027$
\\ \hline\hline  
\multicolumn{4}{|c|}{{\bf Type III Shortcuts }}\\ \hline \hline 
Fig.\ref{T3-braneb}, $v(0)=0.145$     &$1.49$&$0.970$&$0.210$ \\ \hline
Fig.\ref{T3-braneb}, $v(0)=0.140$     &$1.75$&$1.394$&$0.523$ \\
    \hline\hline 
Fig.\ref{T3-braned}, $v(0)=-5\times10^4$   &$0.58$&$353$  &$3.51$  \\ \hline
Fig.\ref{T3-braned}, $v(0)=-10^5$     &$1.11$&$683$  &$26.6$ 
    \\ \hline 
Fig.\ref{T3-braned}, $v(0)=-1.2\times 10^5$&$1.33$&$823$&$48.1$ 
\\ \hline
Fig.\ref{T3-braned}, $v(0)=-1.5\times 10^5$ &$1.69$&$1058$&$104$
    \\ \hline\hline 
Fig.\ref{T3-branef}, $v(0)=-0.8$      &$0.73$&$0.753$&$0.123$
    \\ \hline 
Fig.\ref{T3-branef}, $v(0)=-0.9$      &$0.78$&$0.808$&$0.148$
    \\ \hline 
Fig.\ref{T3-branef}, $v(0)=-1$        &$0.81$&$0.842$&$0.111$ \\ \hline
Fig.\ref{T3-branef}, $v(0)=-1.04$     &$0.80$&$0.831$&$-0.129$
    \\ \hline\hline 
Fig.\ref{T3-braneg}, $v(0)=-5\times 10^4$&$0.61$&$372$&$4.11$
\\ \hline
Fig.\ref{T3-braneg}, $v(0)=-10^5$     &$1.14$&$703$&$29.2$ \\ \hline
Fig.\ref{T3-braneg}, $v(0)=-1.5\times 10^5$&$1.72$&$1081$&$112$ 
\\ \hline
Fig.\ref{T3-braneg}, $v(0)=-2\times 10^5$   &$2.42$&$1568$&$354$
    \\ \hline\hline 
\end{tabular} 
\end{center}
\caption{Domain wall time $\tau$ and time delays $\Delta\tau$ for
  shorcuts appearing in Type II and Type III Solutions} 
\label{sc}
\end{table}

\section{Conclusions}

We have considered here again the question of shortcuts in a Universe 
described by a membrane embeded in a bulk with two extra dimensions, the 
so called dowainwall, described by Einstein gravity with a scalar field. 
In \cite{chre} the full set of solutions of the brane equation of motion 
and Israel conditions at the wall has been obtained. We studied the 
possibility of shortcuts in those cases. In one of our previous works we 
found \cite{Abdalla2} that in a static brane embedded in a pure AdS space 
with a black hole (AdS-Schwarzschild, AdS-RN) shortcuts may appear if a 
certain range of parameters is chosen \cite{abdacasali}. Later, we proved
that in a dynamical brane universe shortcuts are actually quite common
and may provide an alternative explanation to the horizon
problem. 

Indeed, if gravitational shortcuts are common before
nucleosynthesis in a realistic model, they may provide an alternative
to inflation in order to thermalize the early universe. In the present
model, where the Universe is replaced 
by a domain wall, we have proved that shortcuts may exist and above
all abundant, which is a necessary condition in order to solve the
homogeneity problem.

This is not sufficient though and further considerations in a more
realistic setup are certainly needed. We should also
stress that a time-varying dilaton could generate a 
detectable variation of the Newton constant, which imposes certain
constraints on the parameters of the present model.

However, in spite of its shortcomings, the model shows interesting results.
Moreover, we further show that
the delay of the time of flight inside the brane may be comparable
with the time of flight of the graviton itself. This lends further
support for a thermalization via graviton exchange through extra
dimensions, though still not a proof of the solution of the problem.
  
The question to be answered now is whether more realistic models
including e.g. further extra dimensions, such as in the Horava - Witten
formulation, can display similar features.

\bigskip
{\bf Acknowledgements:} This work has been supported by Funda\c c\~ao
de Amparo \`a Pesquisa do Estado de S\~ao Paulo {\bf (FAPESP)} and Conselho
Nacional de Desenvolvimento Cient\'\i fico e Tecnol\'ogico {\bf
(CNPq)}, Brazil. 


\end{document}